\documentclass[manuscript]{acmart}

\AtBeginDocument{%
  \providecommand\BibTeX{{%
    \normalfont B\kern-0.5em{\scshape i\kern-0.25em b}\kern-0.8em\TeX}}}

\copyrightyear{2023}
\acmYear{2023}
\setcopyright{acmlicensed}\acmConference[CHI '23]{Proceedings of the 2023 CHI Conference on Human Factors in Computing Systems}{April 23--28, 2023}{Hamburg, Germany}
\acmBooktitle{Proceedings of the 2023 CHI Conference on Human Factors in Computing Systems (CHI '23), April 23--28, 2023, Hamburg, Germany}
\acmPrice{15.00}
\acmDOI{10.1145/3544548.3581105}
\acmISBN{978-1-4503-9421-5/23/04}

\usepackage{xcolor}
\usepackage{subcaption}
\usepackage{microtype}
\usepackage{array}
\usepackage{enumitem}
\usepackage{colortbl}
\usepackage{graphicx}
\newcolumntype{\$}{>{\global\let\currentrowstyle\relax}}
\newcolumntype{^}{>{\currentrowstyle}}

\def\markup{0}
\if\markup1
\newcommand{\rv}[1]{{\leavevmode\color{blue}#1}}
\else
\newcommand{\rv}[1]{#1}
\fi

\begin{document}

\title{Enhancing Older Adults' Gesture Typing Experience Using the T9 Keyboard on Small Touchscreen Devices}


\author{Emily Kuang}
\affiliation{%
  \institution{Golisano College of Computing and Information Sciences}
  \institution{Rochester Institute of Technology}
  \city{Rochester}
  \state{New York}
  \country{USA}
}
\email{ek8093@rit.edu}

\author{Ruihuan Chen}
\affiliation{%
  \institution{School of Information}
  \institution{Rochester Institute of Technology}
  \city{Rochester}
  \state{New York}
  \country{USA}
}
\email{rc8895@rit.edu}

\author{Mingming Fan}
\authornote{Corresponding Author}
\orcid{0000-0002-0356-4712}
\affiliation{
  \institution{Computational Media and Arts Thrust}
  \institution{The Hong Kong University of Science and Technology (Guangzhou)}
  \city{Guangzhou}
  \country{China}
}
\affiliation{
  \institution{Division of Integrative Systems and Design \& Department of Computer Science and Engineering}
  \institution{The Hong Kong University of Science and Technology}
  \city{Hong Kong SAR}
  \country{China}
}
\email{mingmingfan@ust.hk}

\renewcommand{\shortauthors}{Emily Kuang, Ruihuan Chen, Mingming Fan}

\begin{abstract}
Older adults increasingly adopt small-screen devices, but limited motor dexterity hinders their ability to type effectively. 
While a 9-key (T9) keyboard allocates larger space to each key, it is shared by multiple consecutive letters. Consequently, users must \rv{interrupt} their gestures when typing consecutive letters, leading to inefficiencies and poor user experience. 
Thus, we proposed a novel keyboard that leverages \rv{the} currently unused key 1 to duplicate letters from the previous key, allowing the entry of consecutive letters without interruptions. 
\rv{A user study with 12 older adults showed that it significantly outperformed the T9 with wiggle gesture in typing speed, KSPC, insertion errors, and deletes per word while achieving comparable performance as the conventional T9.}
\rv{Repeating the typing tasks with} 12 young adults found \rv{that} the advantages \rv{of the novel T9 were} consistent \rv{or enhanced}. 
We also provide error analysis and design considerations for improving gesture typing on T9 for older adults.
\end{abstract}



\maketitle

\section{Introduction}

The world is currently experiencing rapid growth in two areas: the aging population and the adoption of technology. 
The World Health Organization predicts that the proportion of the world's population over 60 years old\footnote{\rv{We follow the definition from the United Nations, which considers people over the age of 60 to be ``older adults'' \cite{united_nations_older_2022}.}} is expected to double from 12\% to 22\% between 2015 and 2050 \cite{world_health_organization_ageing_2021}. 
Although the use of technology is lower among older adults as compared to young people, there is still an explosion in their use of computers, smartphones, and other forms of technology \cite{czaja_usability_2019}. 
In the US, 18\% of older adults owned smartphones in 2013, and that number more than doubled to 42\% in just four years \cite{anderson_tech_2017}. 
Furthermore, a survey of 1824 older adults in Switzerland in 2020 showed that up to 6.6\%  owned a smartwatch \cite{seifert_smartwatch_2020}. 
The key services that older adults use on smartwatches were \rv{making phone calls}, keeping track of their health information (e.g., pedometer and heart rate monitor), and getting notifications \cite{fernandez-ardevol_my_2017, manini_perception_2019}. 
To fulfill health-tracking functionalities, smartwatch apps require text input from older adults. 
One such example is the use of smartwatches for real-time collection of pain scores, where participants entered their scores throughout their day \cite{manini_perception_2019}.
However, due to the small screen and age-related declines in motor dexterity, it can be hard for older adults to type accurately and effectively on smartwatch interfaces \cite{hawthorn_designing_2006, nicolau_elderly_2012, voelcker-rehage_motor-skill_2008}.
Moreover, some older adults even felt panicked when sending text messages due to the difficulty of text entry \cite{kurniawan_older_2008}.
Text entry on small-screen devices is inherently difficult since users need to reach a small target using their relatively large fingertips \cite{Lin18, Siek2005fatfinger}. 
The T9 keyboard---a predictive text technology in a $3\times3$ layout---is often adopted on such devices, especially smartwatches \cite{Qin18, komninos_text_2014}. 
Compared to the 26-key QWERTY layout, T9 allocates 3-4 letters on each key, resulting in bigger space for each key and thus alleviates inaccurate input (e.g., due to fat-finger problems) \cite{Siek2005fatfinger}.
Currently, the T9 keyboard can be used in existing commercial devices with both Android WearOS (e.g., KeyOboard \cite{darienzo_keyoboard_2022}) and iOS (e.g, WatchKey \cite{vulcan_labs_company_limited_watchkey_2022}, Type Nine \cite{porsager_type_2022}.  
T9 is also the default keyboard on the Samsung Galaxy Watch 4 \cite{romero_how_2022}. 
Due to its advantages in efficiency and on small devices, there is ongoing work to optimize \rv{the} T9 keyboard (e.g., Optimal-T9 used a computational approach \cite{Qin18}, and SmartVRKey explored T9 usage in a VR environment \cite{adhikary_smartvrkey_2018}).
Furthermore, the T9 keyboard was adopted by feature phones in \rv{the} early 2000s and many older adults are familiar with the layout \cite{fuglerud_studying_2018}.
Thus, we were motivated to investigate ways to improve text entry on \rv{the} T9 keyboard on small-screen devices for older adults.

As touchscreen keyboards gradually replaced physical keypads, many users adopted gesture typing, which allows them to input a word through one continuous movement~\citep{Zhai12}. 
In contrast to tap typing, gesture typing does not require the user to have precise operations on the touch position and has been leveraged for text entry on small touchscreens due to improved error tolerance \citep{Wong2018fingert9} and tactile feedback \cite{Qin18}.
Recent research has shown that gesture typing is particularly promising for older adults. 
In an experiment comparing gesture typing and tap typing on smartphone QWERTY keyboards, older adults were 15\% faster and had a 27\% lower error rate when using gesture typing~\citep{Lin18}. 
We were inspired by this work and sought to investigate gesture typing on the T9 keyboard on smartwatches for older adults. 



\begin{figure}[htbp]
    \centering
    \includegraphics[height=1.3in]{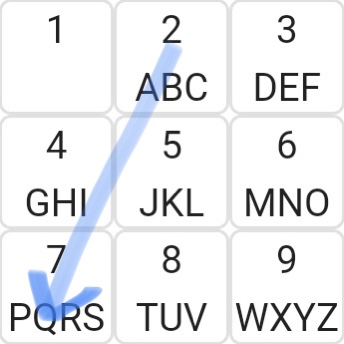}
    \enspace
    \includegraphics[height=1.3in]{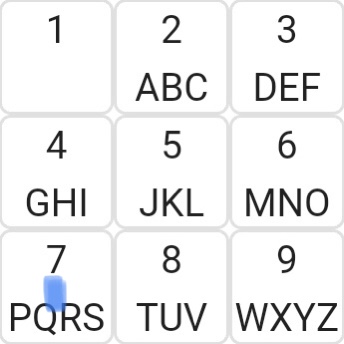}
    \enspace
    \includegraphics[height=1.3in]{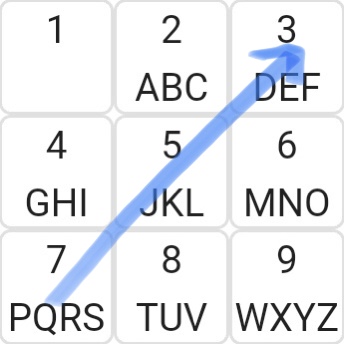}
    \caption{If the user wants to type ``APPLE'' with gestures on the conventional T9 keyboard, they must first swipe from key 2 to key 7, then pause on key 7 or lift their finger up before continuing the direction change to key 3.}
    \Description{A series of 3 annotated images of the T9 keyboard where the blue lines indicate the keystrokes required to type "APPLE"}
\label{fig:typing-apple-t9}
\end{figure}

Despite the individual advantages of using the T9 keyboard or gesture typing, gesture typing directly on the T9 still faces many important challenges.
When gesture typing on the conventional T9 keyboard, the gesture will inevitably be interrupted due to two consecutive letters sharing the same key, leading to typing inefficiencies and poor user experience.
For example, as shown in Fig.~\ref{fig:typing-apple-t9}, when typing the word ``APPLE'', the swipe gesture starts from key 2 to key 7 to type ``AP'', but then the gesture must be interrupted (either with a pause on key 7 or a lift of the fingertip) to re-enter the letter ``P'' before swiping to key 5 and key 3 to complete the word.
Thus, when gesture typing words that contain consecutive letters, the user must perform multiple gestures by lifting the finger several times or pausing for an indefinite period of time without visual or tactile feedback. 
According to the list of frequently used words~\citep{Dolch1936}, 68 out of 220 (30.9\%) frequently used non-nouns and 45 out of 95 (47.4\%) frequently used nouns are affected by this limitation of gesture typing on the T9.
These negative impacts on the typing experience and high frequency of occurrence call for further investigation on how to improve the gesture typing experience on the T9 keyboard for small-screen devices without interrupting the flow of the strokes.


To address this need, we proposed a novel keyboard that enhances gesture typing on the T9 keyboard while maintaining the conventional layout. 
Since key 1 is not currently occupied by any letters, we leveraged this space to duplicate the letters of the previously entered key. 
This allows users to swipe to key 1 from any of the other 8 keys to repeat the same letter without the need to pause on a certain key or lift their finger up.
To evaluate our proposed keyboard, we conducted a user study to compare the T9 with enhanced key 1 with the conventional T9 and the T9 with wiggle gesture proposed by \citet{Bi12a}, which requires users to make three direction changes within the same key to enter a consecutive letter. 
We carried out a within-subjects study with 12 older adults aged 61 to 72 ($Mean = 64.8, SD = 3.72$), who were asked to use all three methods to type 20 phrases randomly selected from Mackenzie's phrase set \citep{Mac2003phrase}.
To see if age makes any difference in the performance and preferences between the three keyboards, we also recruited 12 young adults to complete the same set of typing tasks.
We then compared the three T9 text entry methods in terms of typing performance and subjective feedback.

Our main findings showed that for older adults, the T9 with enhanced key 1 significantly outperformed \rv{the T9 with wiggle gesture in typing speed, keystrokes per character (KSPC), insertion errors, and deletes per word while being on par with the conventional T9.}
The advantages of T9 with enhanced key 1 were more prominent for young adults, leading to significantly better typing speed and KPSC \rv{while reducing insertion errors} over the other two methods. 
Nine (75\%) older adults chose the T9 with enhanced key 1 as their favorite typing method, while all of the young adults did. 
\rv{Although the differences between our novel T9 and the conventional T9 were not significant for older adults, the learning curve and feedback from participants suggest this could be a viable improvement over time.}
The T9 with wiggle gesture was the least preferred for both older and young adults and led to a significantly higher workload (based on NASA-TLX ratings) and inferior typing performance. 
Overall, the approaches proposed in this study demonstrated the feasibility of optimizing the performance and user experience of gesture typing on the T9 keyboard without rearranging its standard layout. 
In sum, we make the following contributions: 
\begin{itemize}
    \item We propose a novel method of gesture typing on the T9 keyboard by harnessing key 1 to duplicate the letters of the previously entered key.
    \item We show the advantages of T9 with enhanced key 1 in terms of performance and subjective ratings through a user study with 12 older adults and 12 young adults.
    \item We provide design considerations for improving gesture typing on the T9 for older adults.
\end{itemize}

\section{Related Work}

\rv{Our study is motivated and informed by prior work in two main areas: text entry for older adults and strategies for optimizing the T9 keyboard.}

\subsection{Text Entry for Older Adults}

\subsubsection{Advantages of Gesture Typing}
\citet{Zhai12} proposed three key factors that affect whether a text input method is accepted by the users, namely input speed, learning cost, and growth. 
In terms of speed, gesture typing offers an advantage over tap typing due to its error tolerance and one-finger operation~\citep{Zhai12}. 
Researchers have investigated gesture typing for older adults and found that it was faster than tap typing, and was very easy for them to learn \citep{Lin18}. 
This may be because the underlying action of sliding a finger from one place to another was commonly seen in touchscreen interaction, which made text entry using gestures feel more natural to older adults \cite{Lin18}.
Gesture typing, in particular, allows users to enter words with rough shapes and placement, which contributes to its error-tolerance properties \citep{Bi12a}. 
This is evident for older adults as lower error rates were observed with gesture typing~\citep{Lin18}. 
Furthermore, when compared with young adults, the gesture accuracy of the older adults did not experience much degradation, showing that it is a promising method for them~\citep{Lin18}. 
These promising results motivated us to further explore how to improve the experience of gesture typing for older adults. 

\subsubsection{Using the T9 Keyboard}
Inspired by the prior work demonstrating the advantages of gesture typing for older adults on the QWERTY keyboard, we sought to investigate gesture typing on the T9 keyboard. 
The T9 layout is especially advantageous for small-screen devices as combining several letters onto one key enlarges the size of each key. 
Increasing the size of the target key making it easier to type according to Fitts' Law \citep{fitts_information_1954}. 
The larger key size is especially important for older adults since they experience age-related degradation in both motor control (with slower and more variable movements) and visual acuity \cite{hawthorn_designing_2006, nicolau_elderly_2012, voelcker-rehage_motor-skill_2008}. 
In fact, keyboards containing only five keys have been developed to support text entry and maximize screen size on mobile devices \cite{dunlop_investigating_2008}. 
However, the T9 provides benefits over such keyboards since it is one of the most well-known multi-letter keyboard layout for users since it was adopted in the early 2000s and is currently available on both Android WearOS and iOS \citep{Wong2018fingert9, fuglerud_studying_2018, darienzo_keyoboard_2022, vulcan_labs_company_limited_watchkey_2022, porsager_type_2022}.
Maintaining a familiar layout may decrease the learning cost, which in turn improves the adoption rate as prior work has shown that a high learning cost negatively impacts the subsequent adoption of input methods for novices~\citep{Zhai12}. 
Thus, we combine the advantages of gesture typing and the large keys on the T9 keyboard in our design to improve the experience and performance of text entry for older adults on small-screen devices. 

\subsection{Optimizing the T9 Keyboard}

\subsubsection{Addressing Ambiguity}
While gesture typing is efficient and error-tolerant, it still exhibits ambiguity when typing words that share a similar or identical gesture~\citep{Alvina16}. 
Especially for multi-letter layouts like the T9, the word collision problem (i.e., words with identical sequences) is a key issue that may result in lower accuracy and speed~\citep{Smith15, Qin18}. 
\citet{Dunlop12} tried to combine the unambiguous property of one-letter layouts like QWERTY with the larger key sizes for multi-letter layouts like the T9. QWERTH, a semi-ambiguous keyboard, was developed to increase the key size and maintain a near-QWERTY layout, but lacked considerations for learnability of the new keyboard~\citep{Dunlop12}.
Similarly, \citet{Smith15} developed a QWERTY-like keyboard that reduced error rates by 52\% and 37\% over the original QWERTY keyboard, but the new method increased the path length of gestures.
While the aim was to optimize gesture typing, their approach involved adjusting the keyboard layout, starting with measures like gesture clarity and QWERTY resemblance~\citep{Smith15}. 
\citet{Smith15} found that tweaking the layout can cause some short-term frustration when participants are first introduced to the modified keyboard, but benefits can arise with long-term usage.

\subsubsection{Addressing Learnability}
Other researchers found similar results that optimizations based on the rearrangement of keys can introduce a learning curve that makes users less willing to adopt the new keyboard~\citep{BiZhai16}. 
To minimize the learning cost, \citet{BiZhai16} followed a rule that only two adjacent keys can be swapped. 
Although it addressed the learnability problem to an extent, it also showed that even swapping one pair of keys can introduce learning cost~\citep{BiZhai16}.  
They found that users still needed to intentionally change their gestures which were already formed into habit based on the original layout.
Specifically for the T9, \citet{Qin18} designed the Optimal-T9 keyboard by introducing Qwerty-bounded constraint (i.e. placing QWERTY’s alphabetical arrangement in a $3 \times 3$ layout) to ensure high learnability. They found that Optimal-T9 outperformed the conventional T9 and other T9-like layouts, while drastically reducing error rate over a 26-key QWERTY keyboard~\citep{Qin18}.
Despite these advantages, the locations of the letters that correspond to each key in the T9 were still modified, which may introduce challenges for older adults who are used to the conventional letter placement.
Across these studies, it is evident that learnability is a significant factor in optimizing typing performance but adjusting the key arrangement will inevitably introduce a steep learning curve for the new layout. 

\subsubsection{Optimizations for Older Adults}
The above optimization approaches were for the general population while other researchers started exploring ability-based optimizations for specific user groups \cite{wobbrock_ability-based_2011, sarcar_ability-based_2018}. 
One such example is a computational approach for improving keyboard designs on touchscreen devices for older adults with cognitive impairments \cite{sarcar_ability-based_2019}.
This optimizer considered parameters that have potential effects for aging users, such as key size, number of keys per row, and number of rows \cite{sarcar_ability-based_2019}. 
After experimenting between 2-10 keys on 1-3 rows, the optimizer selected a $3 \times 3$ grouped keyboard design since the larger sized keys addressed visual and motor deficits \cite{sarcar_ability-based_2019}. 
This is the same layout as the T9, which poses similar advantages for older adults.
However, the arrangement of the letters corresponding to each key was different than any existing keyboards, so learnability remains a challenge for older adults. 
Furthermore, none of the aforementioned studies addressed the issue of the swiping gesture being interrupted by consecutive letters. 
Thus, we see an opportunity to improve the gesture typing experience on the T9 keyboard without modifying the layout of the conventional keyboard nor changing the locations of the letters that correspond to each key.

%

%

\section{Method}

This section describes the design process of the proposed keyboard and the procedure of the evaluation with older and young adults.

\subsection{Keyboard Design}


\subsubsection{T9 with Enhanced Key 1}

The conventional T9 keyboard divides 26 letters between key 2 to key 9. Since key 1 is unused (i.e. did not correspond to any letters), it held the most promise for enhancement. 
In particular, we proposed that key 1 can be used to duplicate the letters from the last entered key. 
Our approach is set apart from previous keyboard optimizations (e.g., \cite{BiZhai16, Qin18, sarcar_ability-based_2018, sarcar_ability-based_2019}) by keeping the same arrangement of keys which effectively harnesses the users’ previous familiarity.
Based on challenges highlighted by prior work, our design objectives were: (1) to allow continuous gestures without interruptions, and (2) to maintain the same layout to improve learnability.
\begin{figure}[ht]
    \centering
    \includegraphics[height=1.3in]{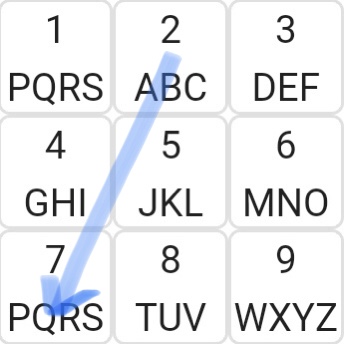}
    \enspace
    \includegraphics[height=1.3in]{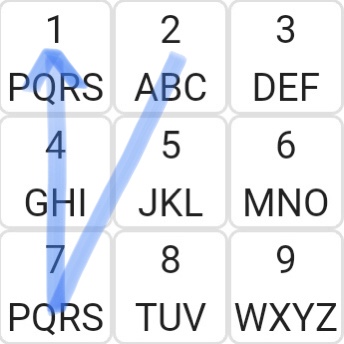}
    \enspace
    \includegraphics[height=1.3in]{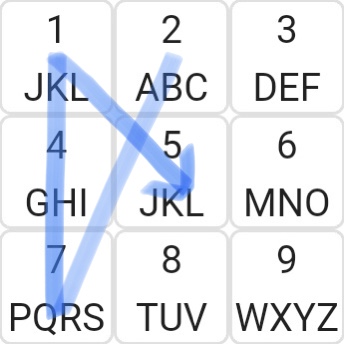}
    \enspace
    \includegraphics[height=1.3in]{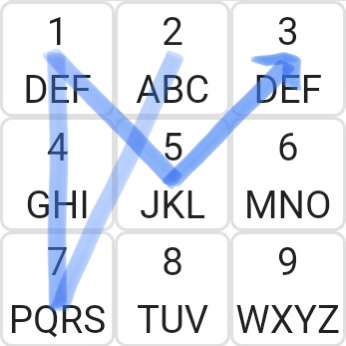}
    \caption{When typing the word ``APPLE,'' the swipe gesture starts from key 2 to key 7 then to key 1, finally to key 5 and key 3. The letters corresponding to key 1 change with the movement of the finger.}
    \Description{A series of 4 annotated images of the T9 with enhanced key 1 where the blue lines indicate the keystrokes required to type "APPLE"}
    \label{fig:typing-apple-t9-enhanced}
\end{figure}

\begin{figure}[ht]
    \begin{center}
    \includegraphics[height=1.3in]{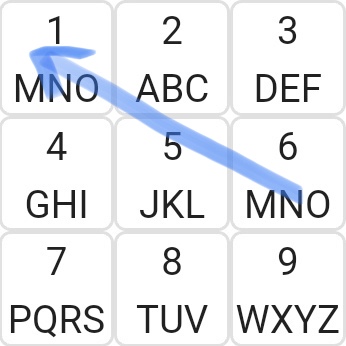}
    \enspace
    \includegraphics[height=1.3in]{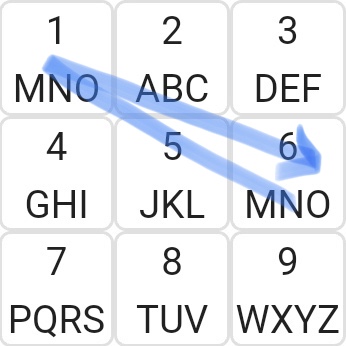}
    \enspace
    \includegraphics[height=1.3in]{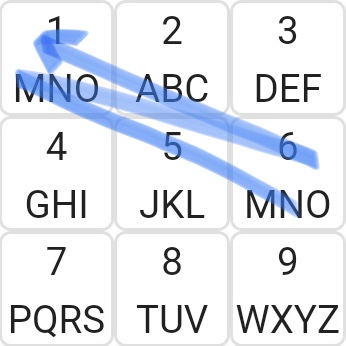}
    \end{center}
    \caption{When typing the word ``MOON'' using the T9 with enhanced key 1 keyboard, the swipe gesture starts from key 6 to key 1, then to key 6, and finally back to key 1.}
    \Description{A series of 3 annotated images of the T9 with enhanced key 1 where the blue lines indicate the keystrokes required to type "MOON"}
    \label{fig:typing-moon-t9-enhanced}
\end{figure}

As shown in Fig.~\ref{fig:typing-apple-t9-enhanced}, when the user enters the word ``APPLE,'' the initial swipe gesture still starts from key 2 and then moves to key 7 for inputting ``AP.''
Then the user can continue the same gesture to key 1 (which duplicates the letters in key 7) to complete the input of the remaining letters ``PLE,'' instead of lifting up their finger to re-enter key 7. 
\rv{For longer sequences of repeated letters such as ``MOON'' (as shown in Fig.~\ref{fig:typing-moon-t9-enhanced}), users can swipe from 6 to 1 to 6 then back to 1 in one single gesture, instead of tapping the same key four times using the conventional T9.}
Additionally, key 1 displays the last-entered key in real-time and acts as a visual indicator of the current status, which follows Nielsen's usability heuristic \cite{nielsen_10_1994}. In contrast with the conventional T9 keyboard where the user's finger may occlude the letters on the key, users can see the letters currently covered by their finger on key 1. 

\subsubsection{T9 with Wiggle Gesture}

\begin{figure}[htbp]
    \centering
    \includegraphics[height=1.3in]{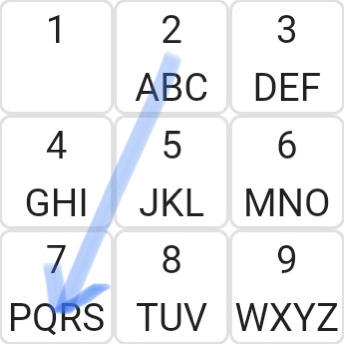}
    \enspace
    \includegraphics[height=1.3in]{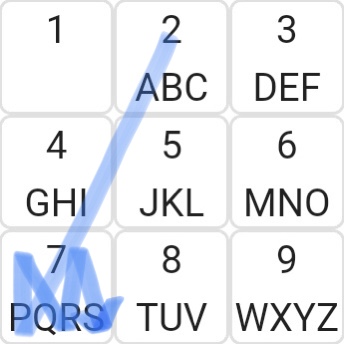}
    \enspace
    \includegraphics[height=1.3in]{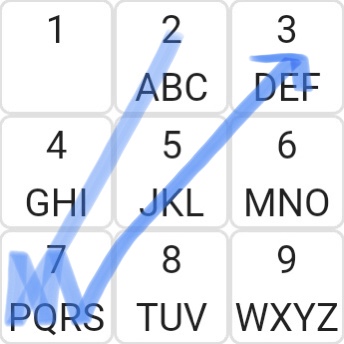}
    \caption{When typing the word ``APPLE'' using the T9 with wiggle gesture, the gesture starts from key 2 to key 7, then a ``W'' or ''M'' is drawn on key 7, and finally to key 3.}
    \Description{A series of 3 annotated images of the T9 with wiggle gesture where the blue lines indicate the keystrokes required to type "APPLE"}
\label{fig:typing-apple-t9-wiggle}
\end{figure}

We extended another gesture typing method proposed by Billah et al., which improves the accessibility of gesture typing and may also be useful for older adults \citep{Bi12a}. 
The wiggle gesture requires the user to draw a wiggly line using their finger within the bounds of a specific key, as shown in Fig.~\ref{fig:typing-apple-t9-wiggle}). 
It is implemented by setting the threshold of the number of swiping direction changes on the X or Y axis to more than three \citep{Bi12a}. 
For example, as shown in Fig.~\ref{fig:typing-apple-t9-wiggle}, for the target word of ``APPLE,'' the user first swipes from key 2 to key 7 to enter ``AP,'' then the user can perform the wiggle gesture on key 7 to re-enter the letter ``P.'' 
Since the wiggle gesture is introduced on every key, this differs from our proposed approach of enhanced key 1 which is a modification on a single key. 
We wanted to explore the potential use of the wiggle gesture since it adheres to our design goals of eliminating interruptions of the gesture when entering consecutive letters while maintaining the same layout to improve learnability.

\subsection{Implementation of the Keyboards}

Due to the COVID-19 pandemic, we implemented the keyboards with the goal of making them easy to distribute so that we could conduct the gesture typing evaluation remotely. 
As such, we developed a web-based evaluation platform that contained the conventional T9, T9 with enhanced key 1, and T9 with wiggle gesture. 
In this way, the participants did not need to install any additional software to access the evaluation platform and the risk of device compatibility issues was reduced. 
The size of the T9 keyboard is 34.8 mm $\times$ 28.6 mm, which is similar to the display area of smartwatches on the market, such as Apple Watch Series 3 \cite{apple_apple_2021}. 
We followed the design criteria for text entry on smartwatches from \citet{dunlop_towards_2014} stating that it should support simple editing (e.g. backspace) and have a reasonably large target (e.g. >7mm). 
In particular, page scaling was disabled for this web application and modern browsers can guarantee similar displays according to the preset CSS configuration. 
While the keyboards simulate the size of a smartwatch, participants were asked to use their own smartphone since it was readily accessible and not all participants owned a smartwatch. 
The application is built using Typescript, Next.js, and Nest.js, and is connected to the MongoDB Atlas database and deployed on Heroku. 
All text entry activities performed on the test platform were recorded by the background program with the timestamp and other metadata, which were synchronized to the database. 

\subsection{Evaluation of Gesture Keyboards}
\label{sec:evaluation_of_gesture_keyboards}
We conducted an IRB-approved online experiment to compare the performance and behavior of older adults using the three keyboards described above. We also conducted the same experiment with young adults to see if the trends \rv{in typing performance} and \rv{subjective} preferences for a certain keyboard are age-specific or consistent across age ranges. 
\rv{We did not specifically compare the typing performance between older and young adults since our goal was to understand whether the patterns between the three keyboards were consistent for different age groups. 
Prior studies have already shown that older adults tend to type slower and make more errors than young adults \cite{Lin18, kalman_writing_2015}.}

\subsubsection{Participants and apparatus}

We recruited 12 older adults (7 females) aged 61 to 72 ($Mean = 64.8, SD = 3.72$) from mailing lists and snowball sampling. 
We also recruited 12 young adults (5 females) aged 20 to 32 ($Mean = 26.1, SD = 3.41$) through posting on the university and research group channels. 
All participants had general professional or native proficiency in English to complete the typing tasks. 
\begin{center}
\begin{table}[tb]
  \caption{The familiarity and years of experience with touchscreen, gesture typing, and T9 keyboards for older and young adults (Five-point scale: 1 - Not at all familiar, 5 - Extremely familiar).}
  \Description{The familiarity and years of experience with touchscreen, gesture typing, and T9 keyboards for older and young adults.}
  \label{tab:pre-survey}
  \begin{tabular}{|p{2.8cm}|l|l|}
    \toprule
     \rowcolor[gray]{0.9} & Older Adults & Young Adults \\
    \midrule
    Familiarity with Typing on Touchscreen & Md=4, IQR=1 & Md=5, IQR=1 \\
    \hline 
    Years of Touchscreen Usage & M=8.17, SD=2.48 & M=9.33, SD=1.72 \\
    \hline
    Familiarity with Gesture Typing & Md=2, IQR=1 & Md=2.5, IQR=1 \\
    \hline
    Years of Gesture Typing Usage & M=1.33, SD=1.87 & M=1.50, SD=2.20 \\
    \hline
    Familiarity with T9 Keyboard & Md=3, IQR=2 & Md=2, IQR=2 \\
    \hline
    Years of T9 Keyboard Usage & M=1.21, SD=2.81 & M=2.25, SD=3.11 \\
  \bottomrule
\end{tabular}
\end{table}
\end{center}
\rv{Table \ref{tab:pre-survey} shows that on average, the older adults were very familiar with typing on the touchscreen, moderately familiar with the T9, but only slightly familiar with gesture typing. 
Young adults were extremely familiar with typing on the touchscreen but less familiar with gesture typing and the T9 keyboard.}

Participants were asked to use their own smartphones to complete the typing tasks. Out of all 24 participants, 14 used iPhones while 10 used Android phones. 
They were also asked to perform gesture typing with the index finger of the dominant hand (22 were right-handed and 2 were left-handed) while they used the other hand to hold the phone. 

\subsubsection{Procedure}

\begin{figure*}[htbp]
    \centering
    \includegraphics[width=\linewidth]{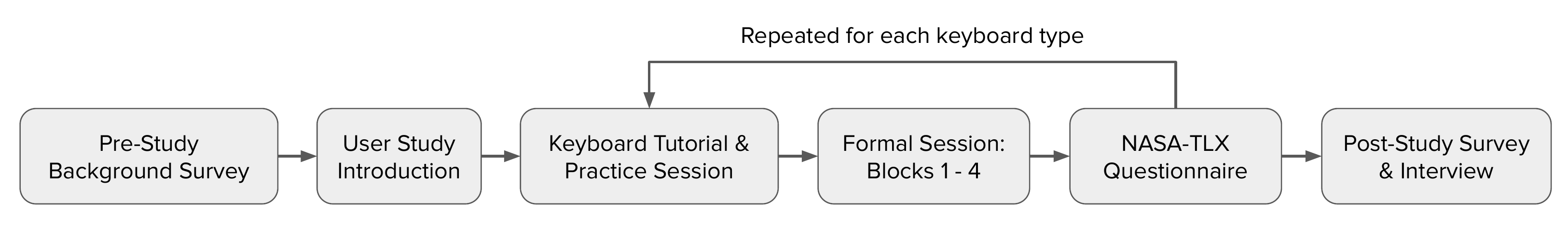}
    \caption{Flowchart of the study procedure followed by each participant.}
    \Description{Flow chart with 6 nodes connected by arrows showing the procedure of the study from the pre-study background survey to the post-study survey and interview}
    \label{fig:procedure}
\end{figure*}

\begin{figure}[htbp]
    \centering
    \includegraphics[width=0.22\textwidth]{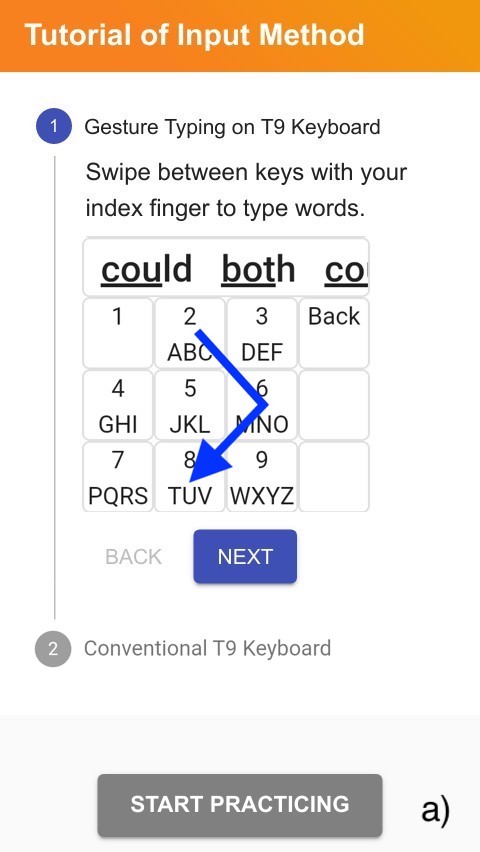}
    \enspace
    \includegraphics[width=0.22\textwidth]{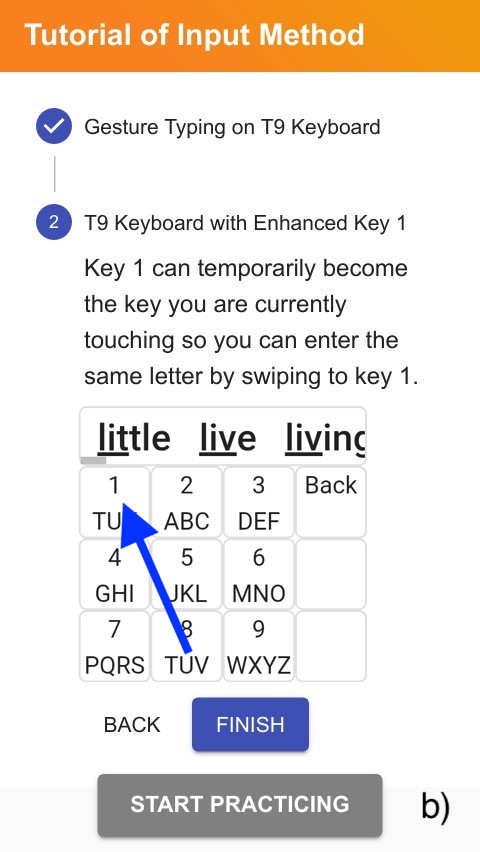}
    \enspace
    \includegraphics[width=0.22\textwidth]{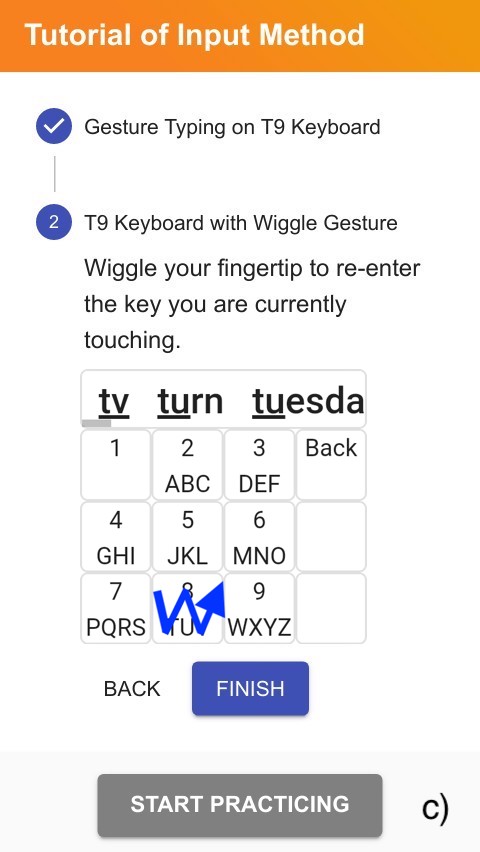}
    \enspace
    \includegraphics[width=0.22\textwidth]{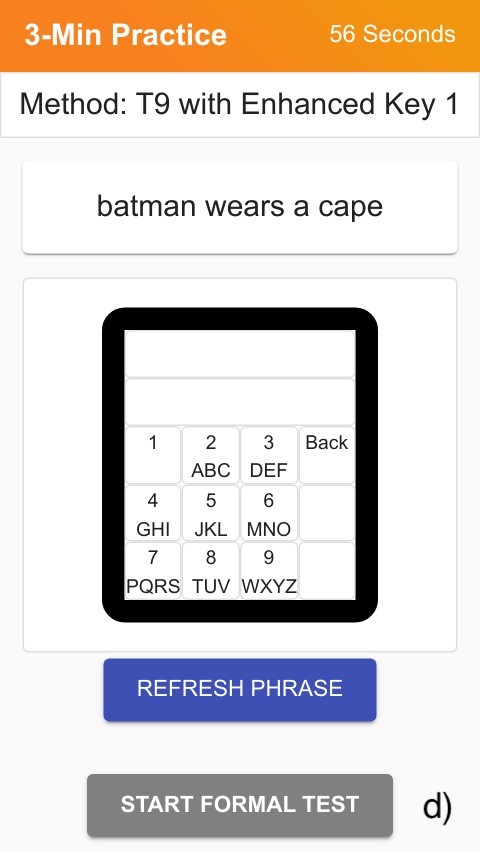}
    \caption{Tutorial for each input method: a) Conventional T9, b) T9 with an enhanced key 1, and c) T9 with wiggle gesture. The blue line indicates the swipe trajectory, with the arrow indicating the direction. d) UI of the practice session, which contains a countdown timer that prevents users from starting the formal tasks until they have become familiar with the keyboard.}
    \Description{A series of 4 screenshots of the evaluation platform where the first 3 screenshots show the tutorial for each input method: a) Conventional T9, b) T9 with an enhanced key 1, and c) T9 with wiggle gesture, where the blue line indicates the swipe trajectory, with the arrow indicating the direction; and the last screenshot shows the UI of the practice session, which contains a countdown timer that prevents users from starting the formal tasks until they have familiarized themselves with the keyboard}
    \label{fig:app-demo}
\end{figure}

The study procedure is shown in Fig.~\ref{fig:procedure}. 
Participants connected to the investigator via Zoom and were given a briefing on the study tasks and instructed to access the link to the test platform with their own smartphone.
Since this study had a within-subject design with the independent variable being the \textit{keyboard type}, all participants were required to complete typing tasks using all three keyboards on the web application as shown in Fig.~\ref{fig:app-demo}. d). 
The order of the type of T9 keyboard was counterbalanced. 
For each keyboard type, participants were first shown a video tutorial of how to input words and then asked to try the keyboard in a 3-minute practice session. They were asked to utilize the special features introduced by two T9 variants to input consecutive letters.

The task for each type of T9 keyboard consisted of 20 phrases randomly selected from Mackenzie's phrase set \citep{Mac2003phrase} divided into 4 blocks, with each block containing 5 phrases. 
Participants were asked to transcribe the target phrase using the keyboard by swiping on the keys and selecting words from the candidate list, which was supported by the unigram language model where the most common words matching a sequence were suggested. 
A space was automatically added after a word was selected from the candidate list since this feature is considered a benefit of predictive keyboards~\citep{Zhai12}.
\rv{For the conventional T9, lifting does not trigger an auto space as it is added only after word selection.}
Participants were also informed that once they selected a word, they could no longer modify it and should continue transcribing the next word 
\rv{, which is in line with prior work \cite{nicolau_elderly_2012}. This design allows us to capture participants’ gesture typing performance instead of their performance with word selection.}
Once participants completed all the typing tasks on a certain keyboard, they were instructed to fill in the NASA-TLX questionnaire before the next keyboard. At the very end, participants filled in a final survey and answered a few questions about the whole experience. The sessions lasted between 1.5 to 2 hours.
In total, our study collected: \textbf{3 keyboards $\times$ 4 blocks $\times$ 5 phrases $\times$ 24 participants $=$ 1440 phrases.} 

\rv{\subsubsection{Data Analysis}
We used the Shapiro-Wilk test to check the normality of all collected data. 
For normal data, we conducted a one-way repeated measures ANOVA to determine whether the means for the three keyboards were significantly different among older adults or among young adults. 
For learnability, we conducted a two-way repeated measures ANOVA with the factors being keyboard type ($k=3$) and block number ($n=4$). 
We also report the effect size with partial eta squared ($\eta_{p}^{2}$) and post-hoc pairwise comparisons with Bonferroni correction, in line with prior typing studies \cite{kalman_writing_2015}. 
For non-normal data that required a non-parametric test, we used the Friedman test for three or more conditions assigned within subjects and pairwise comparisons using Conover's F.
As explained at the beginning of Sec \ref{sec:evaluation_of_gesture_keyboards}, 
 we did not run statistical tests between participant groups (older adults vs. young adults). 
}
\section{Results}

This section contains quantitative findings from the typing tasks which includes typing speed, learnability, keystrokes per character, word error rate, types of errors, and deletes per word, and qualitative findings based on the post-task interviews.

\subsection{Typing Speed (WPM)}

The Words per Minute (WPM) metric is the most frequently used empirical measure of text entry performance and represents the typing speed \cite{hisao_historical_1980, arif_analysis_2009}.
It was calculated using the equation from \citet{Mac2015wpm}, where $S$ is the total number of transcribed characters during the task, and $T$ represents the amount of time taken to transcribe the phrases in minutes. Each five-character string was treated as a single word to convert ``character per minute'' to ``word per minute'' (WPM)~\citep{Mac2015wpm}.

$$
WPM = \frac{|S| - 1}{T} \times \frac{1}{5}
$$

\begin{figure}[htbp]
    \centering
    \includegraphics[width=\linewidth]{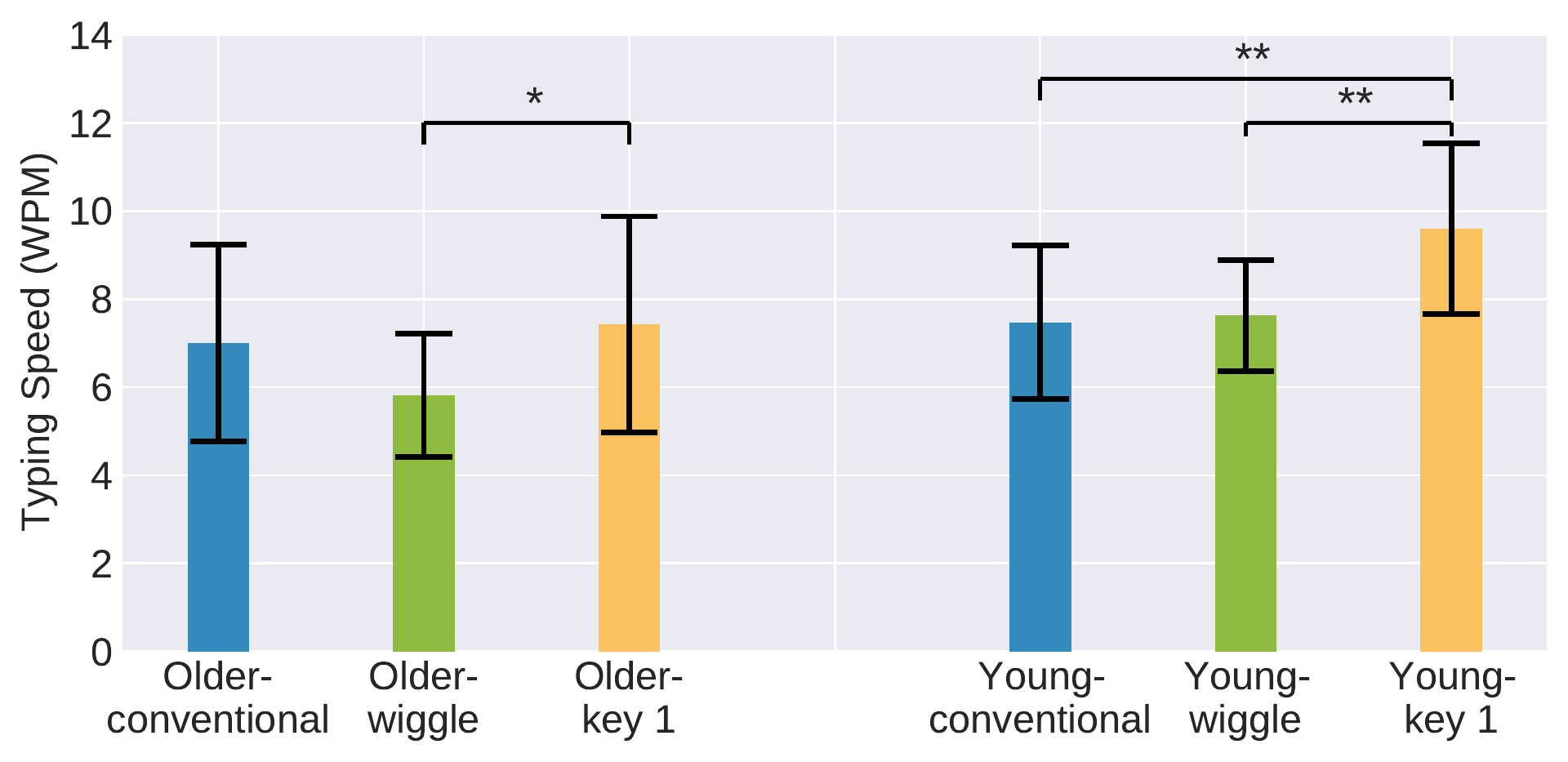}
    \caption{Average typing speed for each T9 keyboard and user group. \rv{(Error-bars show $\pm 1$ SD; * $p < .05$, ** $p < .01$, *** $p < .001$)}}
    \Description{Bar chart showing the average typing speed for each T9 keyboard where T9 with enhanced key 1 resulted in higher speed than the other two keyboards for both older and young adults}
    \label{fig:report-wpm}
\end{figure}

Fig.~\ref{fig:report-wpm} shows the average typing speed of each T9 keyboard for both older and young adults. 
The fastest speed for older adults occurred for T9 with enhanced key 1 which was 7.42 WPM (SD = 2.46), followed by the conventional T9 with 6.85 WPM (SD = 2.37), while the T9 with wiggle gesture was the slowest on average with 5.84 WPM (SD = 1.39).
\rv{ANOVA showed a main effect of keyboards on the typing speed ($F_{2, 22} = 3.79, p < .05, \eta_{p}^{2} = 0.26$).\footnote{\rv{$\eta_{p}^{2} = 0.01$ indicates a small effect, $\eta_{p}^{2} = 0.06$ indicates a medium effect, and $\eta_{p}^{2} = 0.14$ indicates a large effect \cite{spss_effect_2022}.}} Pairwise comparisons revealed a significant difference between T9 with enhanced key 1 and T9 with wiggle gesture ($p < .05$).}
This trend was the same for young adults: the T9 with enhanced key 1 resulted in the fastest typing speed of 9.60 WPM (SD = 1.94). 
\rv{There was also a significant difference due to keyboard type ($F_{2, 22} = 9.69, p < .001, \eta_{p}^{2} = 0.47$), with T9 with enhanced key 1 resulting in faster typing speed than the other two keyboards (both $p < .01$).}

\subsection{Learnability}

\begin{figure}[htbp]
    \centering
    \includegraphics[width=\linewidth]{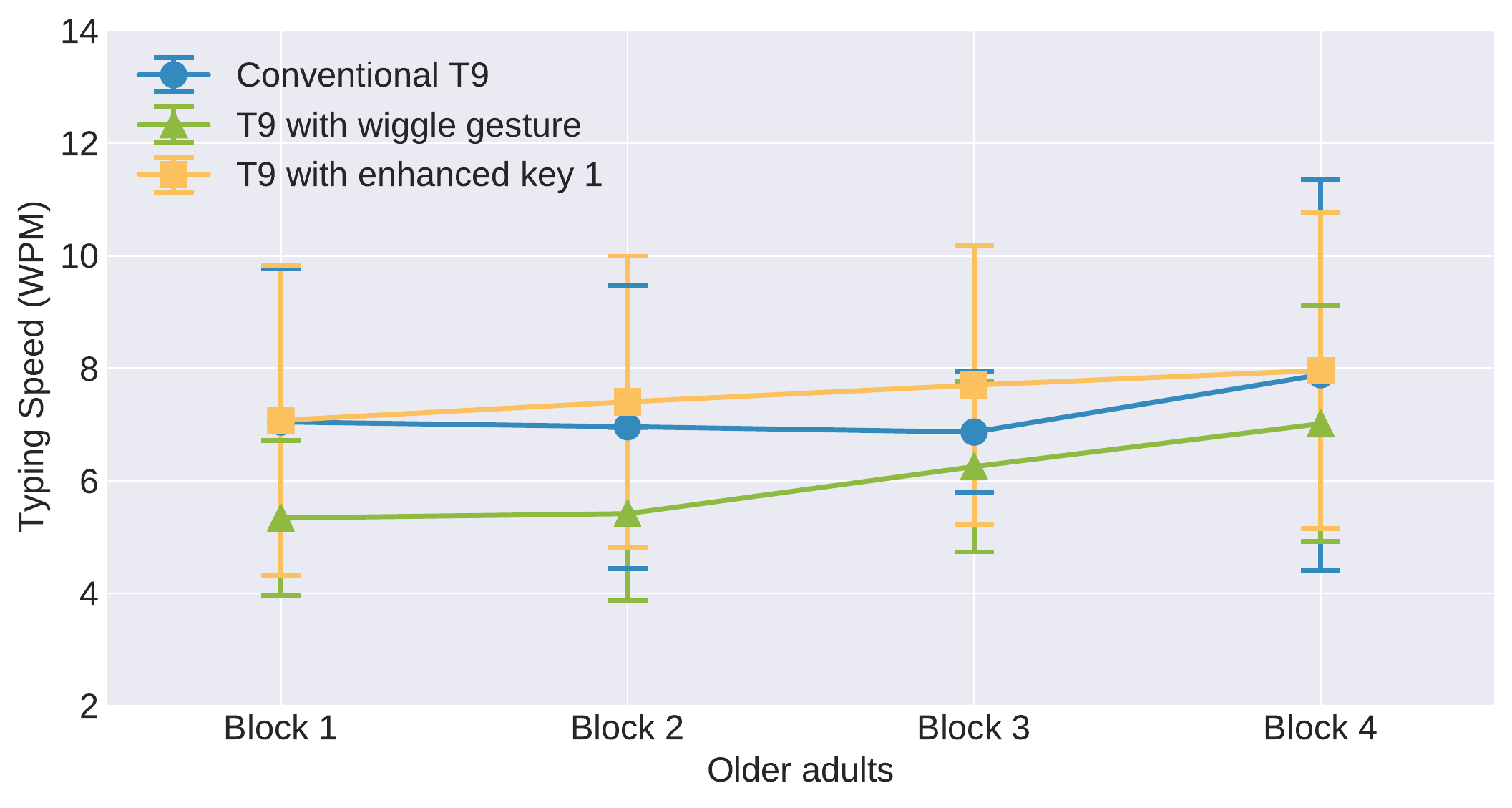}
    \enspace
    \includegraphics[width=\linewidth]{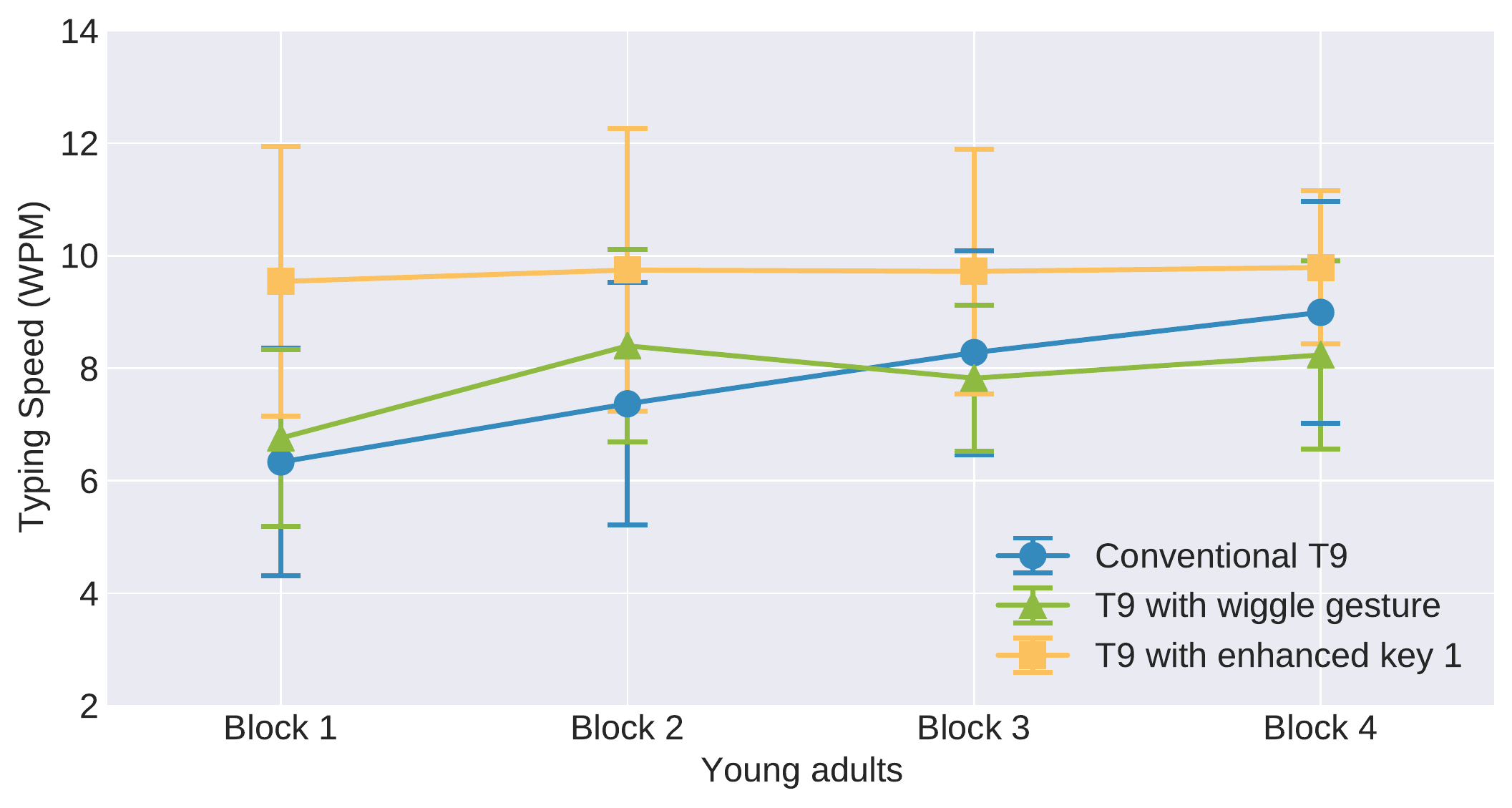}
    \caption{Average typing speed by T9 keyboard and block for older adults (top) and young adults (bottom).}
    \Description{Line chart containing 3 lines for each keyboard type showing the typing speed separated into 4 blocks where there is an overall upward trend showing both older and young adults reaching the highest speed in the fourth block}
    \label{fig:report-learnability}
\end{figure}

Learnability is an important metric for illustrating the learning curve of keyboards.
We used the same calculation as the typing speed (WPM) but separated the measures into blocks. 
The trend for each T9 keyboard across the 4 blocks for older adults is shown in Fig.~\ref{fig:report-learnability} (top). 

\rv{For older adults, the effect of keyboard type on WPM was significant ($F_{2, 22} = 4.24, p < .05, \eta_{p}^{2} = 0.28$), and the effect of block number on WPM was also significant ($F_{3, 33} = 5.65, p < .005, \eta_{p}^{2} = 0.34$). The keyboard $\times$ block interaction effect was not significant.}
The speed for T9 with enhanced key 1 gradually increased throughout each block, which suggests that this could be a viable improvement over a period of use.
For the conventional T9, the speed stayed consistent for the first three blocks before increasing in the fourth block. This may be because participants were already familiar with the conventional T9 and did not experience a steep learning curve at the beginning.
However, for both the T9 with enhanced key 1 and conventional T9, there were no significant differences between any two blocks. 
For the T9 with wiggle gesture, there were significant differences between the first and fourth block ($p < .05$), and the second and fourth block ($p < .05$). 
This suggests that the T9 with wiggle gesture led to the largest learning curve. 

We repeated this analysis for young adults, as shown in the graph on the right of Fig.~\ref{fig:report-learnability}. 
\rv{The effect of keyboard type on WPM was significant ($F_{2, 22} = 8.28, p < .005, \eta_{p}^{2} = 0.43$), and the effect of block number on WPM was also significant ($F_{3, 33} = 12.4, p < .0001, \eta_{p}^{2} = 0.53$). We also observed a significant keyboard $\times$ block interaction effect ($F_{6, 66} = 2.85, p < .05, \eta_{p}^{2} = 0.21$).}
For the T9 with enhanced key 1 keyboard, there were no significant differences between any two blocks as the speed remained fairly consistent at around 10 WPM throughout all 4 blocks. This suggests that young adults did not experience a large learning curve and grasped the method for enhanced key 1 quickly. 
For the conventional T9, young adults were significantly faster in the third block ($p < .05$) and fourth block ($p < .05$) compared to the first, which is shown by the positive slope of the blue line in Fig.~\ref{fig:report-learnability}.
For the T9 with wiggle gesture, the speeds for the second ($p < .05$) and fourth block ($p < .05$) were significantly higher than the first. 

\subsection{Keystrokes Per Character (KSPC)}

\begin{figure}[htbp]
    \centering
    \includegraphics[width=\linewidth]{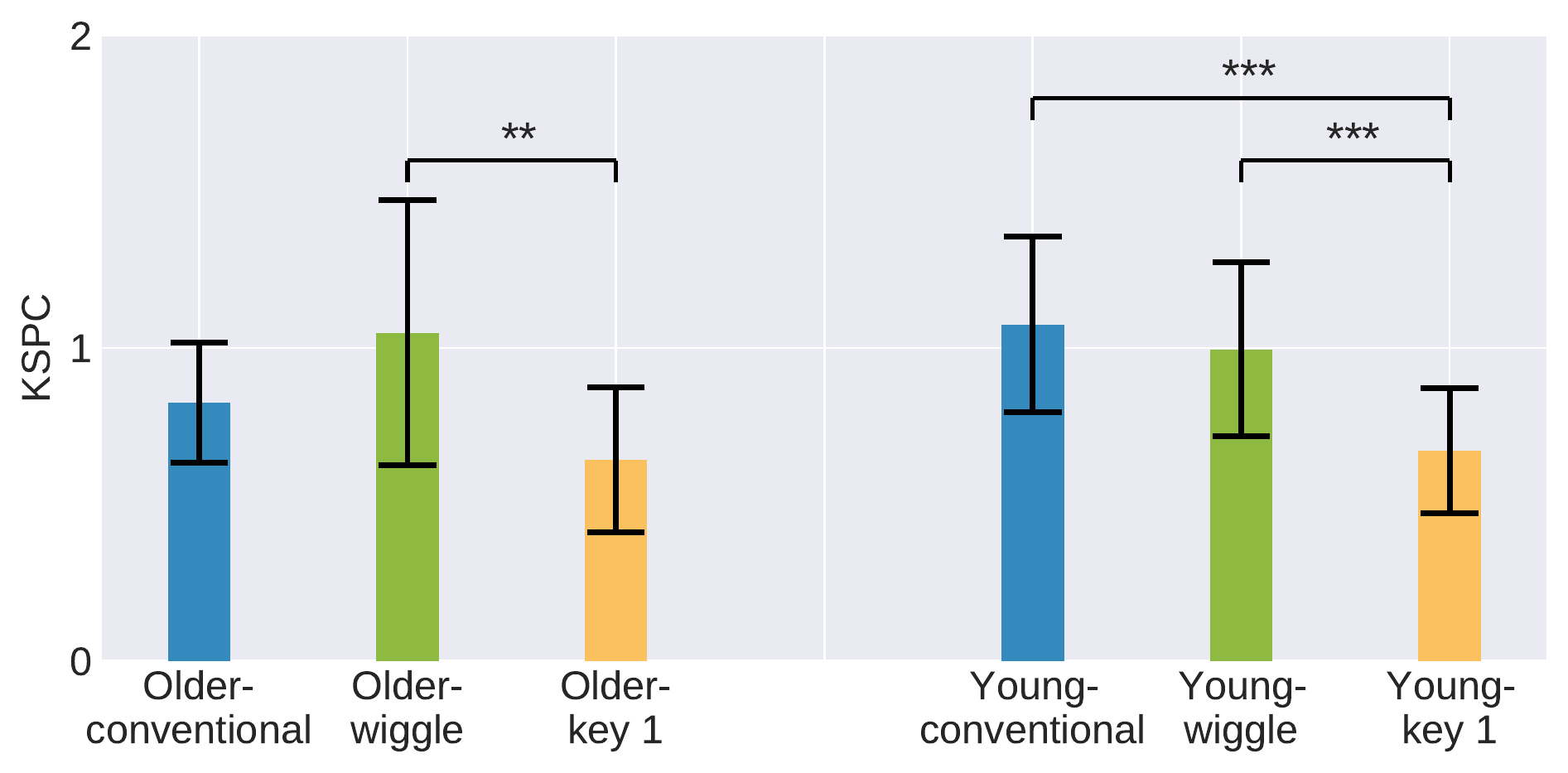}
    \caption{KSPC for each T9 keyboard and user group.}
    \Description{Bar chart showing the KSPC for each T9 keyboard where the T9 with enhanced key 1 led to the lowest keystrokes for both older and young adults}
    \label{fig:report-kspc}
\end{figure}

KSPC is the average number of keystrokes necessary to generate each character~\citep{Mac2002kspc} and can represent the \textbf{efficiency of a keyboard}~\citep{Zhai12}. It was calculated using the following equation: 

$$
KSPC = \frac{Gestures(S) + WordSelections(S) + Deletes(S)}
{CharacterLength(S)}
$$

where $S$ is the transcribed phrase, $Gestures(S)$ is the number of gestures performed to transcribe $S$, $WordSelections(S)$ is the number of times a word was selected from the candidate list, $Deletes(S)$ is the number of character deletes, and $CharacterLength(S)$ is the number of total characters in $S$.
\rv{For example, typing ``APPLE'' without any errors would require 1 gesture using the T9 with enhanced key 1 and T9 with wiggle gesture, but 2 gestures with the conventional T9.}
As shown in Fig.~\ref{fig:report-kspc}, the T9 with enhanced key 1 resulted in the lowest KSPC for older adults. The average KSPC was 0.68 (SD = 0.23) for T9 with an enhanced key 1, 0.88 (SD = 0.20) for conventional T9, and 1.10 (SD = 0.42) for T9 with wiggle gesture. 
Keyboard type had a main effect on the KSPC ($F_{2, 22} = 10.1, p < .001\rv{, \eta_{p}^{2} = 0.48}$) and pairwise comparisons revealed a significant difference between T9 with enhanced key 1 and T9 with wiggle gesture ($p < .01$).
\rv{For young adults,} keyboard type also had a main effect on the KSPC ($F_{2, 22} = 16.6, p < .0001 \rv{, \eta_{p}^{2} = 0.60}$).
\rv{The KSPC for the T9 with enhanced key 1 (0.71, SD = 0.20) was significantly lower than the other two keyboards (both $p < .001$.}

\subsection{Word Error Rate (WER)}

\begin{figure}[htbp]
    \centering
    \includegraphics[width=\linewidth]{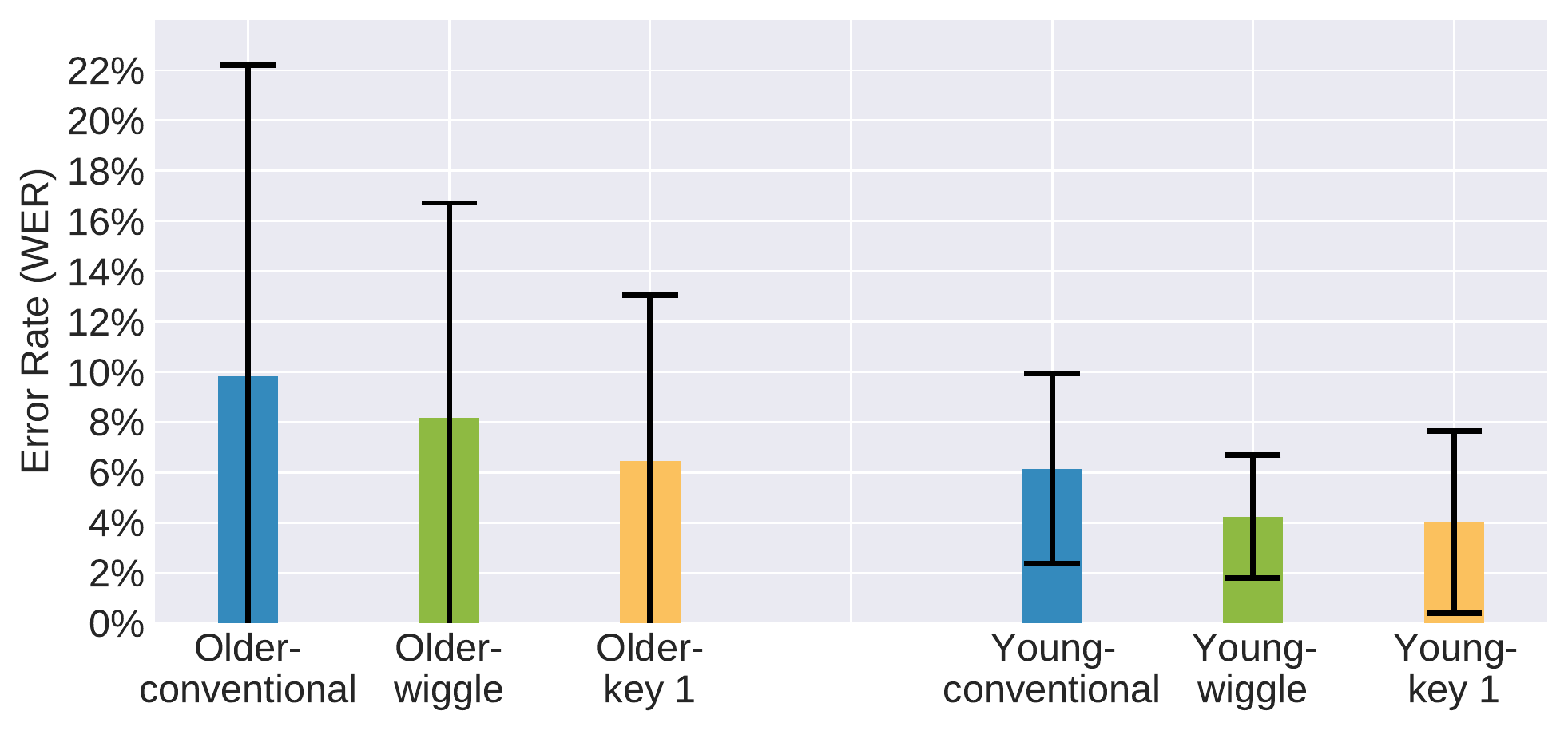}
    \caption{Word error rate for each T9 keyboard and user group.}
    \Description{Bar chart showing the WER for each T9 keyboard where the T9 with enhanced key 1 was the lowest for both older and young adults}
    \label{fig:report-wer}
\end{figure}

Unlike the straightforward calculation of WPM, the error rate is more complex as it is difficult to distinguish errors corrected during entry from those that remain in the transcribed text \cite{arif_analysis_2009, soukoreff_metrics_2003}.
In line with prior work (e.g., \cite{BiZhai16, Qin18}), we used the uncorrected error rate based on the minimum word distance (MWD) between the transcribed phrase $S$ and the target phrase $P$. Specifically, it was calculated using the following equation:

$$
WER = \frac{MWD(S, P)}{WordLength(P)} \times 100\%
$$

where $WordLength(P)$ is the number of words contained in $P$.
As shown in Fig.~\ref{fig:report-wer}, T9 with enhanced key 1 led to the lowest error rate on average for both older adults and young adults. 
The older adult participants made the most input errors on the conventional T9 ($9.82\%, SD = 12.4\%$), followed by the T9 with wiggle gesture ($8.18\%, SD = 8.54\%$), and the least error occurred on the T9 with enhanced key 1 ($6.45\%, SD = 6.61\%$). 
This trend stayed consistent for young adults as well. 
However, \rv{Friedman test} did not find any significant effects of keyboard type on the WER for either older adults or young adults. 

\subsection{Types of Errors}

\begin{figure}[htbp]
    \centering
    \includegraphics[width=\linewidth]{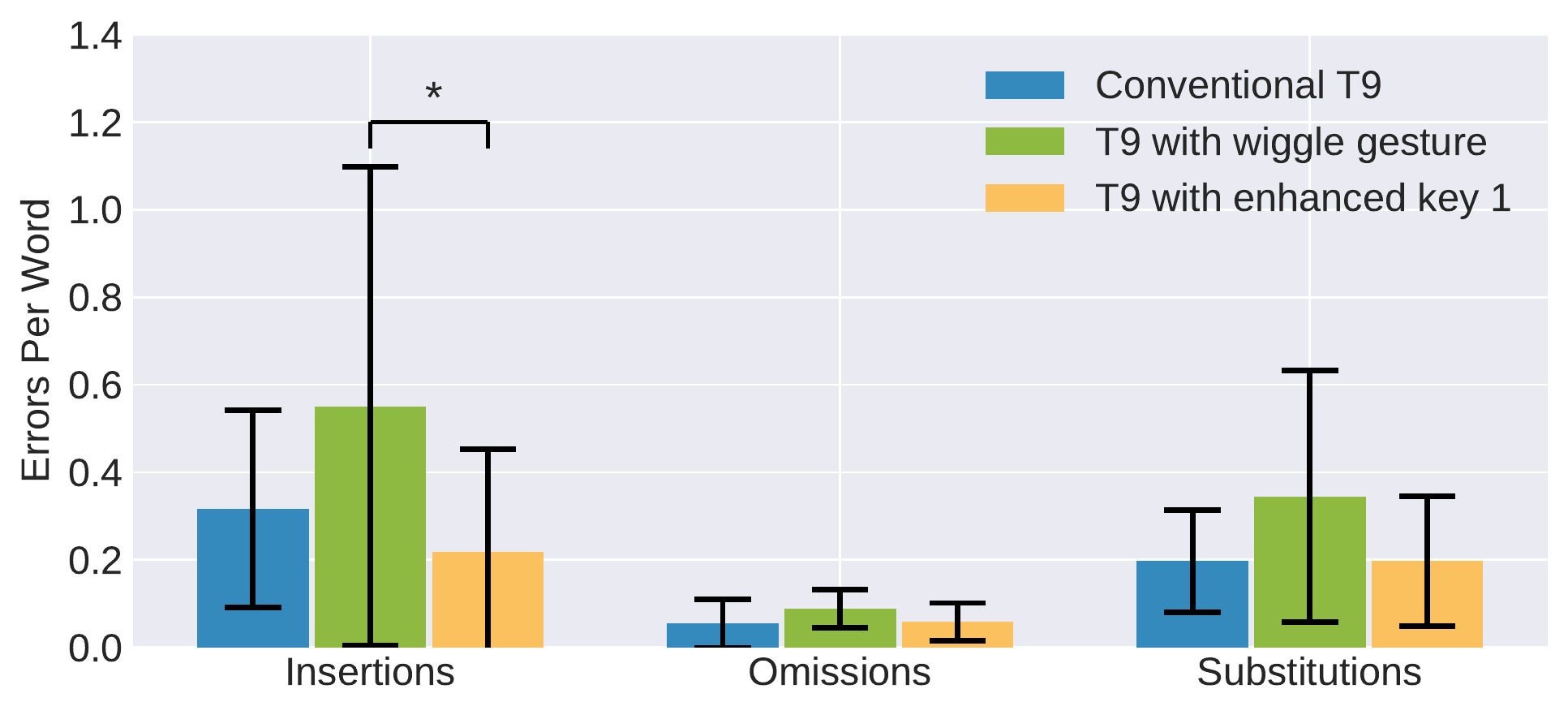}
    \enspace
    \includegraphics[width=\linewidth]{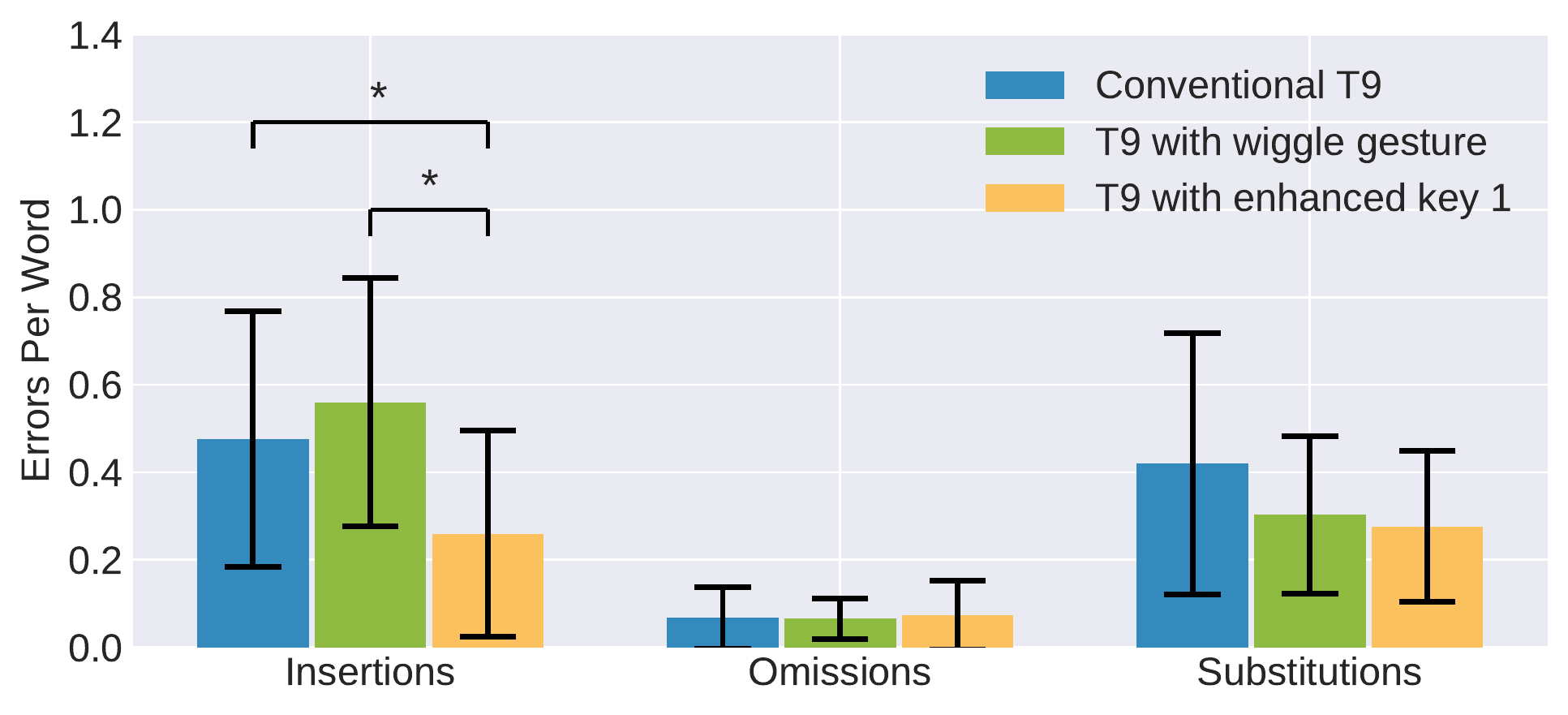}
    \caption{Types of errors made per word for older adults (top) and young adults (bottom).}
    \Description{Two bar graphs showing the number of insertions, omissions, and substitutions per word for older adults on the top and young adults on the bottom separated by keyboard type}
    \label{fig:report-error-type}
\end{figure}

In addition to the word error rate, we conducted keystroke-level analysis on the types of errors that were made by older adults and young adults (as shown in Fig. \ref{fig:report-error-type}). 

There are three common categories of typing errors: \textit{insertion} occurs when an additional key is pressed, \textit{omission} (also called \textit{deletion}) occurs when a key is missed, and \textit{substitution} occurs when an incorrect key is pressed instead of the target key \cite{wobbrock_analyzing_2006}. 
Since T9 only has 9 keys, we analyzed keystroke sequences corresponding to the number of the entered key. For example, the target sequence for ``APPLE'' is [2, 7, 7, 5, 3].
The special entry of consecutive letters using the T9 with enhanced key 1 and T9 with wiggle gesture was denoted by '1,' (e.g. ``APPLE'' is represented by [2, 7, 1, 5, 3]). 
To account for this, we replaced all occurrences of ``1'' with the previous key so that we could compare them with the target sequence. 
Since the T9 is a predictive keyboard, users can select the target word from the candidate list after only partially entering it. Therefore, all entered sequences that were a subset and appeared in the same order as the target sequence was deemed error-free. 

For older adults, we found that the most common error was insertions ($52.2\%$ of all errors), followed by substitutions ($35.1\%$), then omissions ($12.7\%$). 
\rv{Friedman test found a significant difference for insertions ($\chi^{2} = 7.167, p < .05$), but not substitutions or omissions.}
Older adults made significantly more insertions with the T9 with wiggle gesture than the T9 with enhanced key 1 ($p < .05$). 
The distribution of errors was similar for young adults, where the most common were insertions ($50.9\%$ of all errors), followed by substitutions ($40.6\%$), then omissions ($8.5\%$). 
\rv{Similarly, the only significant difference in keyboard type was for insertions ($\chi^{2} = 8.043, p < .05$), which was lower for enhanced key 1 than the other two ($p < .05$).}

\subsection{Deletes Per Word}

\begin{figure}[htbp]
    \centering
    \includegraphics[width=\linewidth]{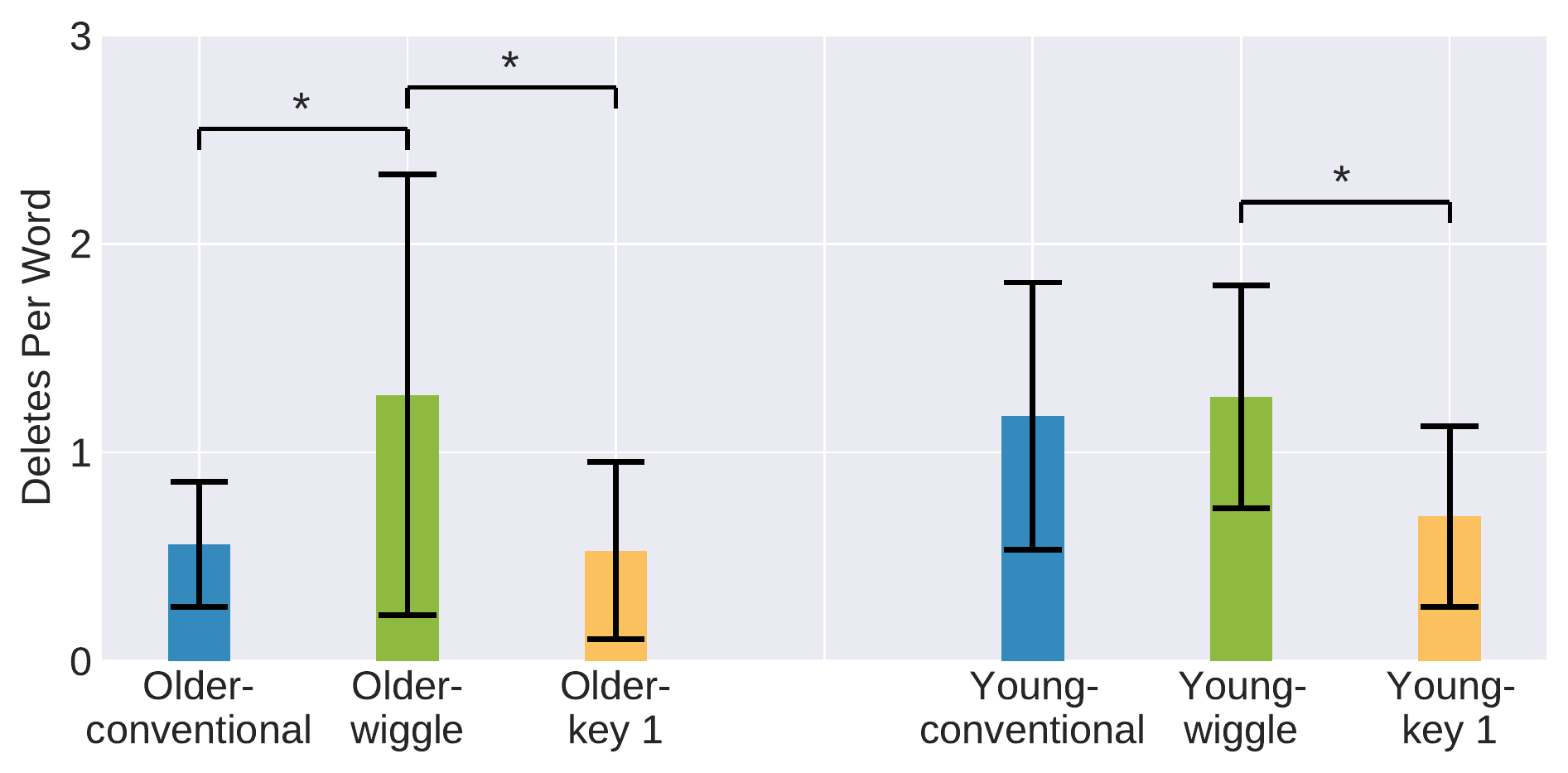}
    \caption{Deletes per word for each T9 keyboard and user group.}
    \Description{Bar chart showing the deletes per word for each T9 keyboard where the T9 with enhanced key 1 led to the lowest number of deletes for both older and young adults}
    \label{fig:report-deletes}
\end{figure}

Deletes per word was calculated to compare the performance in terms of corrected input errors~\citep{Qin18, Palin19, Lin18}. 
As seen in Fig.~\ref{fig:report-deletes}, the T9 with an enhanced key 1 resulted in the least amount of backspace usage for both older and young adults. 
Older adults used on average 0.53 deletes (SD = 0.42) on the T9 with enhanced key 1, 0.57 deletes (SD = 0.30) for conventional T9, and 1.27 deletes (SD = 1.06) for T9 with wiggle gesture. 
There was a main effect of keyboard type on the average deletes per word ($F_{2, 22} = 7.65, p < .005 \rv{, \eta_{p}^{2} = 0.41}$),
\rv{where the T9 with enhanced key 1 resulted in a significantly lower number of deletes than T9 with wiggle gesture ($p < .05$).}
\rv{Similarly, there was a main effect of keyboard type for young adults ($F_{2, 22} = 7.33, p < .005$), where the difference was significant between the T9 with enhanced key 1 and T9 with wiggle gesture ($p < .05$)}.





\subsection{Subjective Ratings}

\begin{figure}[htbp]
    \centering
    \includegraphics[width=\linewidth]{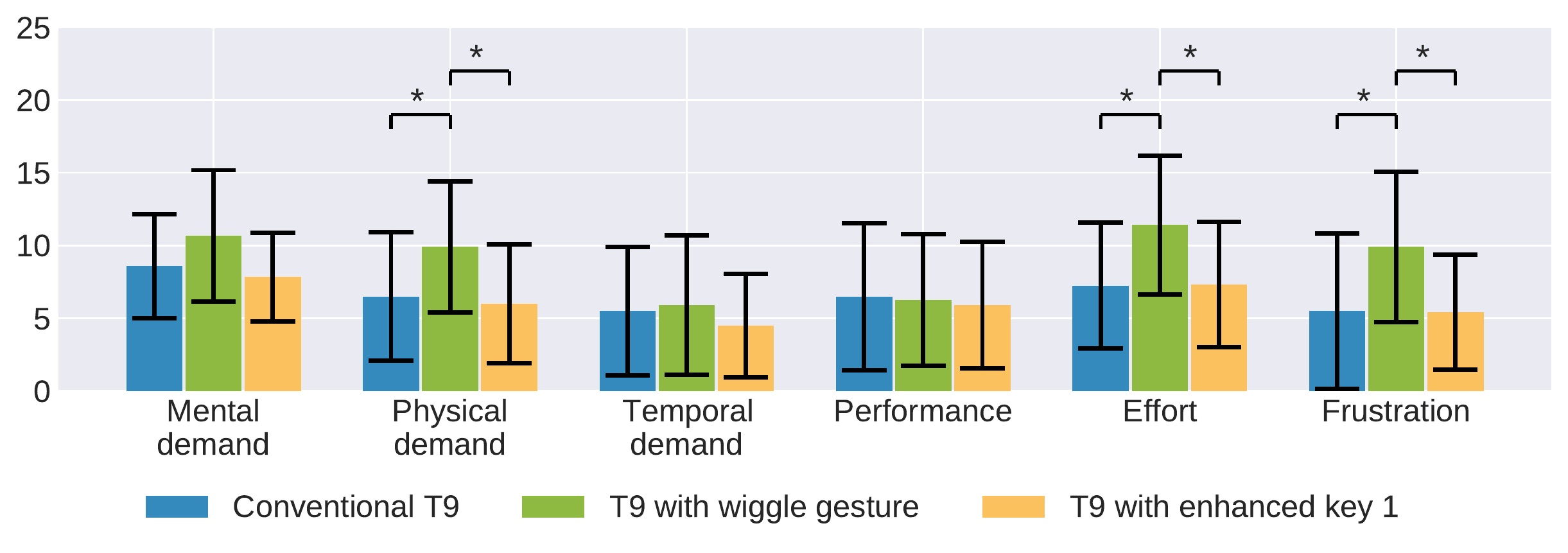}
    \enspace
    \includegraphics[width=\linewidth]{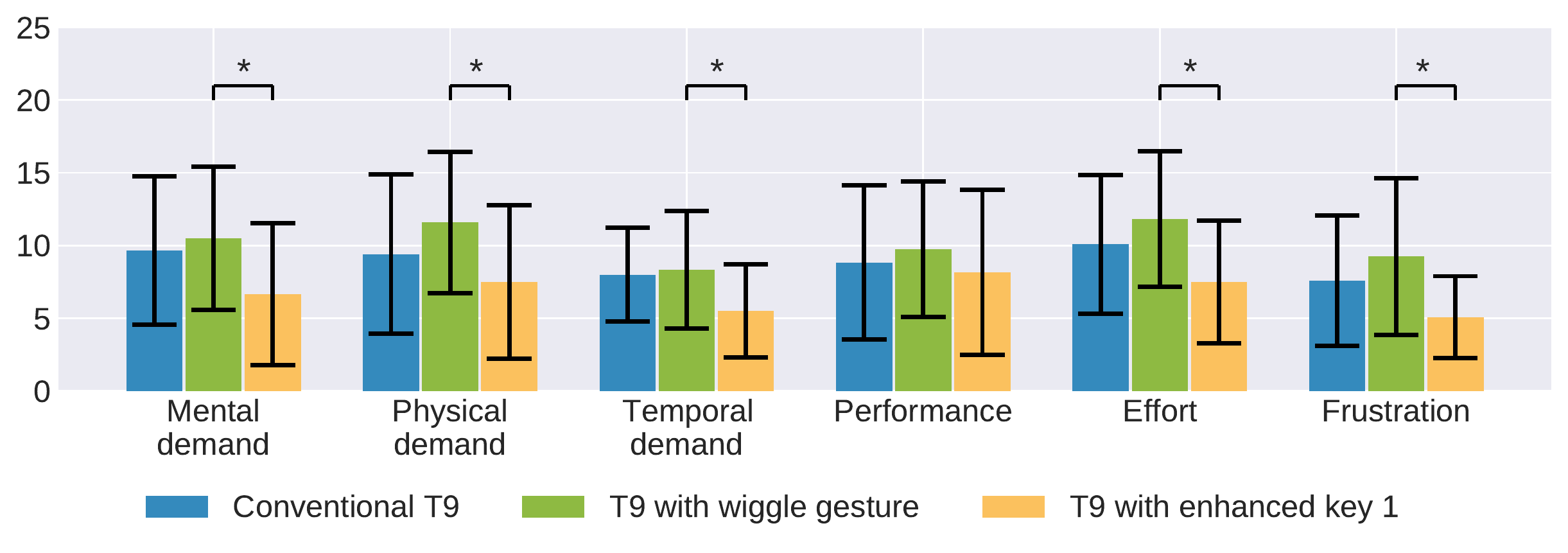}
    \caption{Average ratings on the NASA-TLX of three T9 keyboards by older adults \rv{(top) and younger adults (bottom)} (1: the most positive rating and 20: the most negative rating).}
    \Description{Bar chart of older adults' and young adults' ratings with 3 bars representing the 3 keyboards divided into 6 categories of the NASA-TLX scale where T9 with enhanced key 1 led to the lowest rating for all 6 categories}
    \label{fig:report-nasa-tlx}
\end{figure}

After completing all transcription tasks for a given keyboard, participants filled in the NASA-TLX scale \citep{Hart1988nasa}.
Fig.~\ref{fig:report-nasa-tlx} shows the subjective ratings for older adults \rv{on the top and those of young adults on the bottom.}
\rv{For older adults, the T9 with enhanced key 1 was rated as incurring significantly lower physical demand ($F_{2, 22} = 13.2, p < .0005, \eta_{p}^{2} = 0.55$), effort ($F_{2, 22} = 8.30, p < .005, \eta_{p}^{2} = 0.43$), and frustration level ($F_{2, 22} = 9.78, p < .001, \eta_{p}^{2} = 0.47$) than the T9 with wiggle gesture (all $p < .05$). There were no significant differences between the conventional T9 and T9 with enhanced key 1.
The trend was similar for young adults, with the T9 with enhanced key 1 rated as requiring lower mental demand ($F_{2, 22} = 6.69, p < .01, \eta_{p}^{2} = 0.38$), physical demand ($F_{2, 22} = 6.04, p < .01, \eta_{p}^{2} = 0.35$), temporal demand ($F_{2, 22} = 5.09, p < .05, \eta_{p}^{2} = 0.32$), effort ($F_{2, 22} = 4.07, p < .05, \eta_{p}^{2} = 0.27$), and frustration level ($F_{2, 22} = 5.72, p < .05, \eta_{p}^{2} = 0.34$) than the T9 with wiggle gesture (all $p < .05$).}



\begin{figure}[htbp]
    \centering
    \includegraphics[width=\linewidth]{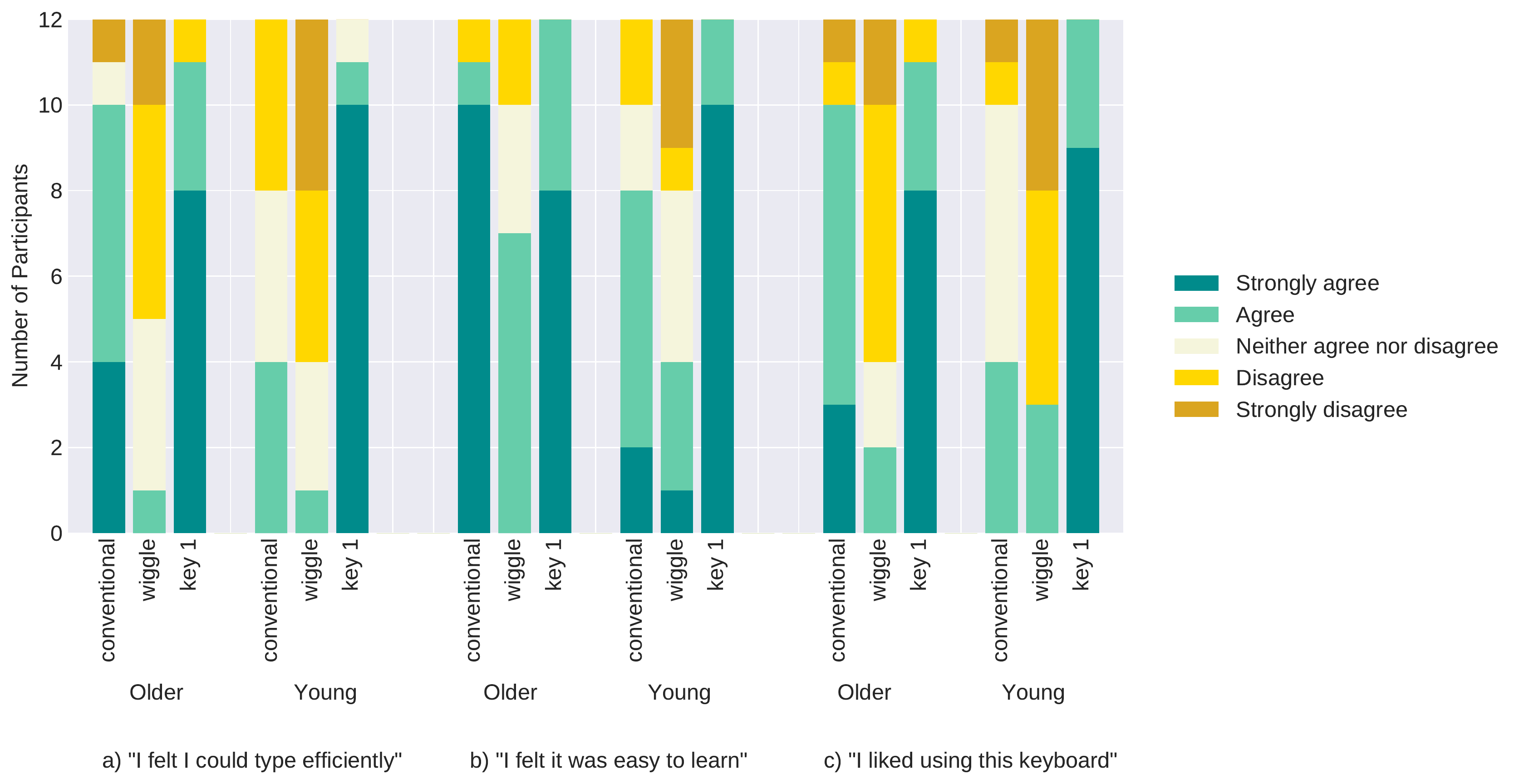}
    \caption{Distribution of the Likert scale responses to whether participants could type efficiently, felt it was easy to learn, and liked using each keyboard.}
    \Description{Stacked bar graph on a divergent color scale showing the distribution of the Likert scale responses to whether participants a) could type efficiently, b) felt it was easy to learn, or c) liked using each keyboard}
    \label{fig:report-likert}
\end{figure}

In addition to filling in the NASA-TLX, participants also responded to three Likert scale questions. Fig.~\ref{fig:report-likert}. a) shows the distribution of their responses to the phrase ``I felt that I could type efficiently'' using each keyboard.
\rv{Older adults rated the T9 with enhanced key 1 as most efficient ($Md = 5, IQR = 1$), followed by conventional T9 ($Md = 4, IQR = 1$), then T9 with wiggle gesture ($Md = 2, IQR =1$).
There was a main effect for keyboard type ($F_{2, 22} = 27.3, p < .0001, \eta_{p}^{2} = 0.71$), where both conventional T9 ($p < .01$) and T9 with enhanced key 1 ($p < .01$) were rated as significantly more efficient than the T9 with wiggle gesture.
The trend was the same for young adults ($F_{2,22} = 36.5, p < .0001, \eta_{p}^{2} = 0.77$), but T9 with enhanced key 1 was considered significantly more efficient than both the conventional T9 ($p < .01$) and T9 with wiggle gesture ($p < .01$).}

Fig.~\ref{fig:report-likert}. b) shows the distribution of their responses to the phrase ``I felt that it was easy to learn how to use the [conventional T9/T9 with enhanced key 1/T9 with wiggle gesture].'' 
\rv{Older adults gave the same median ratings of 5 for conventional T9 ($IQR=0$) and T9 with enhanced key 1 ($IQR=1$), and both were rated as significantly easier to learn than the T9 with wiggle gesture ($F_{2, 22} = 22.3, p < .0001, \eta_{p}^{2} = 0.67$).}
\rv{In contrast, young adults gave the highest rating to the T9 with enhanced key 1 ($Md = 5, IQR = 0$), and considered it significantly easier to learn than both the conventional T9 and T9 with wiggle gesture ($F_{2, 22} = 16.2, p < .0001, \eta_{p}^{2} = 0.59)$.}


Lastly, participants provided ratings on their overall preference for each keyboard, as shown in Fig.~\ref{fig:report-likert}. c). 
\rv{Older adults preferred T9 with enhanced key 1 the most ($Md = 5, IQR = 1$), followed by conventional T9 ($Md = 4, IQR = 0.25$), then T9 with wiggle gesture ($Md = 2, IQR = 1$).}
ANOVA confirmed that there was a main effect of keyboard type ($F_{2, 22} = 22.5, p < .0001\rv{, \eta_{p}^{2} = 0.67}$). Specifically, both conventional T9 ($p < .01$) and T9 with enhanced key 1 ($p < .01$) were preferred over the T9 with wiggle gesture, but there were no significant differences between the two. 
\rv{The trend was the same for young adults ($F_{2,22} = 29.2, p < .0001\rv{, \eta_{p}^{2} = 0.72}$), but pairwise comparisons revealed that the T9 with enhanced key 1 was significantly preferred over the other two keyboards (both $p < .01$).}

\subsection{Qualitative Feedback}

\rv{During the post-task interviews, we asked participants what they liked and what challenges they encountered with each keyboard, as well as the reasons behind their preference ratings. The responses were transcribed and coded independently by two researchers before consolidating.}
To differentiate between groups of participants, older adult participants are denoted ``O1-12'' while young adult participants are denoted ``Y1-12'' in the following sections. 

\subsubsection{Conventional T9}

For the conventional T9, three older adult participants mentioned that they preferred this keyboard over the two variants. In contrast, none of the young adults preferred this keyboard and three of them mentioned that this was their least favorite. This difference may be due to older adults being more familiar with the conventional keyboard since they have used it in the past, while the new variants required more time and effort to learn. This was mentioned by O5 who said \textit{``I'm more used to the conventional T9 so it was the easiest for me to use.''}

When asked about the disadvantages of the keyboard, some older adults mentioned that once they lifted their finger up, sometimes they forgot to continue gesture typing and started tap typing instead (O3).
For typing consecutive keys, Y1 noted that it was easy to tap on the adjacent key after lifting his finger up. When entering words that contain multiple consecutive letters such as ``sprawling'' (where the first 3 letters are all on key 7), Y3 mentioned that it was tedious to tap multiple times on the same key and it \textit{``breaks the smooth experience of gesture typing.''} 
Based on the above feedback, the experience of typing consecutive letters on the conventional T9 has room for improvement.

\subsubsection{T9 with wiggle gesture}

Nine \rv{older} adult participants and nine \rv{young} adult participants ranked the T9 with wiggle gesture as their least preferred keyboard in the study. 
The explanations for this ranking were similar for both user groups. 
Participants mentioned three main pain points: (1) there was \textbf{insufficient feedback} as to how much ``wiggling'' they had to perform, (2) the keyboard size was \textbf{too small} to change the swiping direction, and (3) the wiggle gesture was \textbf{prone to errors}. 
Regarding the first pain point, while participants knew that a wiggle gesture required three direction changes, they still found it hard to keep track of how many changes they performed and how many were actually detected by the keyboard.
Y6 suggested that adding an additional signal or pattern that shows the number of direction changes in real-time (such as a number indicator or progress bar) would give the user better control and more visual feedback. 
In terms of the small size, this was especially evident for older adults where many participants mentioned that it was hard to wiggle inside the same key because the region was too small to make those direction changes. 
Oftentimes, they accidentally touched the adjacent key when wiggling and unintentionally entered the wrong character, which resulted in more time to press backspace and re-enter the desired character again. 
For the third pain point, many participants mentioned that it was easy to overcompensate for the wiggle gesture and mistakenly enter two consecutive letters instead of just one. 
Regarding learnability, many participants felt that the wiggling gesture required more practice than the other keyboards, \rv{which is} evident in the subjective ratings for the ease of learning shown in Fig.~\ref{fig:report-likert}. b). 

\subsubsection{T9 with enhanced key 1}

\rv{Nine older adults chose the T9 with enhanced key 1 as their favorite keyboard while all twelve young adults preferred it.}
When asked about why they liked the enhanced key 1, multiple participants commented on the \textit{``smooth and continuous typing experience,''} (i.e. they didn't have to lift their fingers to break the gesture when typing consecutive letters). 
For example, Y1 said ``I like the enhanced key 1 the most because it gives you the option to continue your gesture, so you don't have to pick up your finger.''
O12 also mentioned that swiping to key 1 offered the smoothest typing experience and she felt it was the fastest to use after getting used to it. 
Another advantage mentioned by Y7 was that key 1 acted as a visual indicator which alleviates \rv{the occlusion of} the current key. 
Furthermore, most participants stated that it was easy to learn this keyboard after practicing it a few times. For the older adults (O4, O5, and O7) who didn't rank this keyboard as their most preferred, they mentioned that they didn't find it difficult to use but were more simply comfortable using the conventional T9 since they had prior experience. Other older adults agreed that they found this method easy to learn, which is supported by the increasing typing speed throughout each block (shown in Fig.~\ref{fig:report-learnability}) and their ratings for ease of learning ($Md = 5, IQR = 1$).

\section{Discussion}

\subsection{Typing Performance}

\subsubsection{Typing Speed}
For older adult participants, we found that T9 with enhanced key 1 led to 6.0\% and 27.5\% faster-typing speed over the conventional T9 and T9 with wiggle gesture respectively, \rv{but the difference was only significant between the enhanced key 1 and wiggle gesture.}
In contrast, the speed advantage of T9 with enhanced key 1 was more apparent for young adults since it was significantly faster than both the conventional T9 by 28.5\% and the T9 with wiggle gesture by 25.8\%. 
\rv{After conducting the study, we found that the resulting speeds are slower than those reported by prior literature. 
For example, participants using WatchWriter achieved average gesture speeds of 24 WPM on a 1.3'' circular display \cite{Gordon_watchwriter_2016}, while participants using VelociTap reached 34.9 WPM on a 25 mm $\times$ 16 mm keyboard \cite{Vertanen_velocitap_2015}. 
However, other studies reported slower speeds: participants using ZoomBoard achieved 9.3 WPM on a 16 $\times$ 6mm keyboard \cite{oney_zoomboard_2013}, while a later study comparing ZoomBoard with Callout and ZShift on a 28.4 $\times$ 11.4 mm keyboard reported 8.2 WPM, 8.3 WPM, and 9.1 WPM respectively. 
While our keyboard is slightly larger (34.8 $\times$ 28.6 mm), the young adult participants achieved comparable speeds on our novel T9 (9.6 WPM) \cite{leiva_text_2015}.
It is important to note that the aforementioned keyboards all employed a QWERTY layout with 26 keys while our study utilized the 9-key layout.
T9 keyboards tend to have slower typing speeds due to the ambiguity of the multi-letter layout. 
Even the optimized-T9 keyboard proposed by previous researchers resulted in a slower typing speed than the conventional QWERTY keyboard \cite{Qin18}. 
Furthermore, prior work added novel features (e.g., statistical decoding, error correction, space key omission, and iterative zooming) with the goal of achieving the best typing performance \cite{Gordon_watchwriter_2016, Vertanen_velocitap_2015, Qin18, oney_zoomboard_2013, leiva_text_2015, leiva_error_2015}.
In contrast, our goal was not to design the fastest keyboards by rearranging layouts or introducing advanced features but to leverage older adults' familiarity with the T9 keyboard toward a smoother typing experience.
We learned that our novel keyboard had an advantage over the T9 with wiggle gesture, but did not significantly improve typing speed over the conventional T9. 
}

\subsubsection{Efficiency \& Error Rate}
In terms of typing efficiency (as indicated by KSPC), T9 with enhanced key 1 resulted in significantly lower keystrokes than T9 with wiggle gesture by 38.2\%. 
\rv{It also had 22.7\% lower keystrokes than the conventional T9 but the difference was not significant.}
Surprisingly, the KSPC of T9 with wiggle gesture was higher than the conventional T9 even though the wiggle gesture was designed to reduce KSPC by minimizing the number of gesture interruptions. 
To investigate this phenomenon, we examined the error rate. 
The average WER for older adults on the T9 with enhanced key 1 was 34.3\% lower than conventional T9 and 20.9\% lower than T9 with wiggle gesture \rv{, although the differences were not significant. 
This trend is similar to another study in which the proposed optimal-T9 resulted in lower WER than the conventional T9 and QWERTY keyboards, but the results were not significant \cite{Qin18}.
The WER of the T9 with enhanced key 1 (6.45\%, SD = 6.61\%) was slightly lower than the reported WER of another study in which older adults used gesture typing on the QWERTY keyboard (6.79\%, SD = 8.53\%) \cite{Lin18}.
Both WERs of the T9 with wiggle gesture and the conventional T9 were higher than the previous study. 
}
In this study, older adults were slightly more accurate when using the T9 with wiggle gesture than the conventional T9, so the extra keystrokes observed may be attributed to the use of deletes to correct typing errors. 
Indeed, the deletes per word for older adults were significantly higher for T9 with wiggle gesture than both conventional T9 and T9 with enhanced key 1. 
The frequent use of deletes demonstrates that older adults experienced considerable difficulty entering letters using the T9 with wiggle gesture.
\rv{Overall, the T9 with wiggle gesture resulted in significantly poorer performance in typing speed, KSPC, and deletes per word, and was rated as requiring significantly more physical demand, effort, and frustration. 
The reasons were revealed in the qualitative feedback as participants felt it was difficult to perform multiple direction changes within a small space without accidentally swiping onto an adjacent key. 
Future improvements should consider enlarging the small key to a comfortable size like ZoomBoard \cite{oney_zoomboard_2013} and removing the adjacent keys so that users can perform the wiggle gesture in a larger area.}

\subsubsection{Types of Errors}
When analyzing the three error types, we found that over half ($52.2\%$) of all errors made by older adults were insertions, followed by substitutions ($35.1\%$) and omissions ($12.7\%$). 
Insertions were the most common due to the prevalence of \textbf{accidental touches} since the keyboard size was very small compared to the participants' fingers. 
Indeed, a prior study comparing tablets vs. mobile phones showed that older adults made significantly more insertion errors with the smaller device \cite{nicolau_elderly_2012}. 
Surprisingly, the types of errors occurred in the opposite order as the results from a study conducted in 2012 with older adults typing on a touchscreen QWERTY keyboard, where they found that the most common types of errors were omission, followed by substitutions and insertions \cite{nicolau_elderly_2012}. 
Since omissions are usually considered cognitive errors that do not depend on motor abilities \cite{kristensson_five_2009, beaver_characterising_2019}, one possible explanation could be that the previous participants were older ($M=79, SD=7.3$) compared to those in our experiment ($M=65, SD=3.7$).
However, neither study accounted for cognitive differences in the participants so this can not be confirmed. 
Another explanation for the difference could be due to the keyboard type. Since the T9 is a predictive keyboard, a space was automatically appended once the participant selected a word from the candidate list. 
In contrast, the prior study required older adults to manually press ``space'' and found that older adults often forgot to enter a blank space between words, which accounted for 25-30\% of all errors \cite{nicolau_elderly_2012}. 
When compared to other user groups such as motor-impaired and non-impaired participants typing with a physical computer keyboard \cite{kane_truekeys_2008}, our results followed the same trend with insertions being the most common and omissions the least common. 
Based on these varying results, further research is warranted to compare the typing errors made by older adults on different types of keyboards.

\subsection{Design Considerations for Improving Gesture Typing on the T9 for Older Adults}

\subsubsection{Maximize keyboard size}
Our study also revealed some challenges that older adults experienced while gesture typing on the T9. 
We found that despite employing the T9 keyboard with larger keys and following the guidelines for keys being at least 7mm \cite{dunlop_towards_2014}, the ``fat finger'' problem still occurred. 
Many of the older adult participants tried to enlarge the keyboard since it only took up a portion of the smartphone screen. 
After informing them that page scaling was disabled to simulate the size of a smartwatch, they mentioned that the keys were still too small relative to their fingertips. 
Thus, future designers should maximize the keyboard space on smartwatches by removing any interface elements that are not vital to the typing task \rv{, as well as using zooming or callout techniques to make the keys more visible \cite{leiva_text_2015}.}
Another possible way to alleviate this issue is to use a stylus which has a smaller tip to enter letters, but this requires the use of additional equipment and may be inconvenient for a smartwatch.  
Future work is warranted to develop a set of guidelines for text entry on small-screen devices specifically for older adults and optimize the interface design. 


\rv{
\subsubsection{Balance the tradeoff between maintaining conventional layouts and optimizing typing speed.}
Based on the learnability curves for T9 with enhanced key 1, the typing speed of both older and young adults increased throughout each block (Fig. \ref{fig:report-learnability}). 
Although we could not conduct a longitudinal typing study due to the practical constraints in participant recruitment, the curves suggest that the T9 with enhanced key 1 could be a viable improvement over a period of use, which supports our design principle of maintaining conventional layouts. 
However, the typing speeds observed in our study are lower than in recent work with other layouts that employed the 26-key QWERTY keyboard \cite{Gordon_watchwriter_2016, Vertanen_velocitap_2015} or a T9-like keyboard with rearranged letters in each key \cite{Qin18}. 
To confirm the benefit of conventional layouts, future work could be done to compare the novel T9 (which maintains the same letters in each key) and the standard QWERTY keyboard. 
Furthermore, future keyboard designers need to balance the learning cost of new layouts with the slower speed of conventional layouts. 
}

\subsubsection{Provide advanced prediction techniques}
We also found that it was considerably challenging for older adults to enter words with long gestures. 
Some words in the phrase set \cite{Mac2003phrase} were particularly difficult to type such as ``sprawling'' where the first three letters were all on key 7. 
Other words like ``anniversary'' and ``subdivisions'' contain 11 and 12 letters respectively, which made it difficult for older adults to select all the keys correctly in one continuous gesture. 
We also received some feedback that target phrases with uncommon or unfamiliar words such as ``racketball'' and ''chlorine'' were more difficult to type. 
The participants needed to refer back to the target phrase and switch their attention between the target and the keyboard while typing to ensure their spelling is correct. 
This suggests that the typing performance of older adults is impacted not only by the keyboard design but also their familiarity with words and length of words. 
Thus, older adults may benefit from \textbf{more advanced techniques that can reduce the number of keystrokes}, such as prediction algorithms that predict words based on partial word input (e.g. text suggestions \cite{microsoft_enable_2022}) or based on patterns of common words that follow the previous word (e.g. Smart Compose \cite{google_use_2022} and next word prediction \cite{stremmel_pretraining_2021}).

\rv{
\subsubsection{Extend the interaction area beyond the touchscreen.}
This study focused on improving the T9 keyboard on small touchscreen devices for older adults. 
However, the issue of ``fat fingers'' on small interfaces is generalizable to all keyboard types as well as target selection tasks.
The size restrictions of smartwatches require new designs and input techniques \cite{funk_using_2014}.
When designing for older adults to use smartwatches, we may consider other ways to augment and extend the interaction interface beyond the touchscreen. 
For example, an elicitation study produced a taxonomy of gestures for 31 smartwatch tasks that included above-device air swipes, rim taps, and hovers \cite{arefin_shimon_exploring_2016}.
Other researchers have shown the viability of using a touch-sensitive wristband for selection and scrolling since the wristband exhibits has a larger surface area than the watc screen \cite{perrault_watchit_2013}.  Researchers then leveraged the wristband for text entry, achieving 2.91 WPM and 3.45 WPM using linear and multitap keyboards respectively \cite{funk_using_2014}. 
As augmented reality (AR) technology becomes more prevalent, researchers have explored the use of virtual keyboards \cite{maiti_preventing_2017}. 
In the future, it may be possible to utilize virtual keyboards for text input, which would no longer confine users to the small touchscreen of a smartwatch. 
}

\section{Limitations and Future Work}
Though this work has produced a novel T9 keyboard with an enhancement on key 1 that was shown to surpass the conventional T9 and T9 with wiggle gesture \rv{in some typing metrics and subjective ratings}, there are areas beyond the scope of this work that warrant further research. 
First, our proposed design focused on addressing the issue of entering consecutive keys on ambiguous keyboards. 
There are other potential bottlenecks when using ambiguous keyboards on small-screen devices. 
Namely, the need to select candidate words from the list generated by the word prediction algorithm can be time-consuming and subject to the ``fat finger'' problem.
Many older adult participants mentioned that they tapped on the wrong word in the candidate list after they had entered the correct gesture sequence, which was another source of frustration. 
Thus, future work can explore the optimization of predictive text algorithms and different interface designs for the candidate list to minimize mistakes in its selection. 
\rv{Furthermore, this study focused on text entry with letters only so only 9 keys were provided. Other layouts such as the 3 $\times$ 4 keyboard with an empty 0 allow the entry of punctuation marks and the full range of numbers (0 - 9). Future work could explore how to leverage the empty 0 to facilitate text entry with punctuation marks and numbers.}

This study employed the most fundamental unigram language model and did not employ features such as auto-completion and auto-correction. 
In practice, increases in the processing power of small-screen devices like smartwatches allow more advanced auto-correct and prediction algorithms to be harnessed. 
However, the potential effects on older adults have not been fully explored. 
Past research suggested that older adults tend to dislike text-prediction algorithms, but this study was completed over 10 years ago \cite{kurniawan_older_2008}. 
Advances in text-prediction technology in the past decade may have increased the acceptance of such algorithms in the older population, and future work can be done to confirm this hypothesis further. 

Some older adult participants regularly use a stylus for text entry on their smartphones due to the ``fat finger'' issue. However, to ensure a valid comparison across user groups and keyboards, they were asked to only use their fingertips for this study. 
Future studies may conduct larger and more thorough empirical investigations of T9 keyboards that span different input strategies such as fingertip and stylus and investigate how different input devices affect the typing performance of older adults. 

\rv{Lastly, there were a low number of participants in our study, which may have resulted in the non-significant differences between our novel T9 and the conventional T9. 
While we did not observe significant differences for every measure, this does not mean that no differences existed. 
To better understand whether a trend holds, one study is unlikely sufficient even with more participants \cite{hornbaek_is_2014}.  Future work should conduct more studies with older adults of different backgrounds to validate our findings and examine how various factors might affect the findings.} 

\section{Conclusion}

Overall our main finding is encouraging for improving the gesture typing experience of older adults on the T9 keyboard. 
Our proposed design utilized key 1 to allow users to make continuous gestures when typing consecutive letters without rearranging the keyboard layout, leading to a better typing experience and learnability. 
We also provided empirical evidence comparing the proposed design with the conventional T9 and T9 with wiggle gesture. 
Through user studies with 12 older adults, we found that the T9 with enhanced key 1 led to \rv{27.5\% faster-typing speed over the T9 with wiggle gesture and was comparable to the conventional T9 (6.0\% faster but not statistically significant).} 
We found that the most common typing errors were insertions ($52.2\%$), substitutions ($35.1\%$), and omissions ($12.7\%$).
\rv{By having young adults complete the same typing tasks, we confirmed that the trends in favor of the T9 with enhanced key 1 either stayed consistent or were more prominent across age groups.}
Finally, we also provided suggestions for future designers and areas of further investigation for improving gesture typing and T9 keyboard layouts for older adults.





\bibliographystyle{ACM-Reference-Format}
\bibliography{main.bib}


\begin{thebibliography}{64}


\ifx \showCODEN    \undefined \def \showCODEN     #1{\unskip}     \fi
\ifx \showDOI      \undefined \def \showDOI       #1{#1}\fi
\ifx \showISBNx    \undefined \def \showISBNx     #1{\unskip}     \fi
\ifx \showISBNxiii \undefined \def \showISBNxiii  #1{\unskip}     \fi
\ifx \showISSN     \undefined \def \showISSN      #1{\unskip}     \fi
\ifx \showLCCN     \undefined \def \showLCCN      #1{\unskip}     \fi
\ifx \shownote     \undefined \def \shownote      #1{#1}          \fi
\ifx \showarticletitle \undefined \def \showarticletitle #1{#1}   \fi
\ifx \showURL      \undefined \def \showURL       {\relax}        \fi
\providecommand\bibfield[2]{#2}
\providecommand\bibinfo[2]{#2}
\providecommand\natexlab[1]{#1}
\providecommand\showeprint[2][]{arXiv:#2}

\bibitem[\protect\citeauthoryear{Adhikary}{Adhikary}{2018}]%
        {adhikary_smartvrkey_2018}
\bibfield{author}{\bibinfo{person}{Jiban Adhikary}.}
  \bibinfo{year}{2018}\natexlab{}.
\newblock \showarticletitle{{SmartVRKey} - A Smartphone Based Text Entry in
  Virtual Reality with T9 Text Prediction}.
\newblock \bibinfo{journal}{\emph{Michigan Technological University}}
  \bibinfo{volume}{{CS}5760 - Human-Computer Interaction \& Usability}
  (\bibinfo{year}{2018}), \bibinfo{pages}{12}.
\newblock


\bibitem[\protect\citeauthoryear{Alvina, Malloch, and Mackay}{Alvina
  et~al\mbox{.}}{2016}]%
        {Alvina16}
\bibfield{author}{\bibinfo{person}{Jessalyn Alvina}, \bibinfo{person}{Joseph
  Malloch}, {and} \bibinfo{person}{Wendy~E. Mackay}.}
  \bibinfo{year}{2016}\natexlab{}.
\newblock \showarticletitle{Expressive Keyboards: Enriching Gesture-Typing on
  Mobile Devices}. In \bibinfo{booktitle}{\emph{Proceedings of the 29th Annual
  Symposium on User Interface Software and Technology}} (Tokyo, Japan)
  \emph{(\bibinfo{series}{UIST '16})}. \bibinfo{publisher}{Association for
  Computing Machinery}, \bibinfo{address}{New York, NY, USA},
  \bibinfo{pages}{583–593}.
\newblock
\showISBNx{9781450341899}
\urldef\tempurl%
\url{https://doi.org/10.1145/2984511.2984560}
\showDOI{\tempurl}


\bibitem[\protect\citeauthoryear{Anderson and Perrin}{Anderson and
  Perrin}{2017}]%
        {anderson_tech_2017}
\bibfield{author}{\bibinfo{person}{Monica Anderson} {and}
  \bibinfo{person}{Andrew Perrin}.} \bibinfo{year}{2017}\natexlab{}.
\newblock \bibinfo{title}{Tech {Adoption} {Climbs} {Among} {Older}
  {Americans}}.
\newblock
\newblock
\urldef\tempurl%
\url{https://www.pewresearch.org/internet/2017/05/17/tech-adoption-climbs-among-older-adults/}
\showURL{%
\tempurl}


\bibitem[\protect\citeauthoryear{Apple}{Apple}{2021}]%
        {apple_apple_2021}
\bibfield{author}{\bibinfo{person}{Apple}.} \bibinfo{year}{2021}\natexlab{}.
\newblock \bibinfo{title}{Apple {Watch} {Series} 3 - {Technical}
  {Specifications}}.
\newblock
\newblock
\urldef\tempurl%
\url{https://support.apple.com/kb/sp766?locale=en_US}
\showURL{%
\tempurl}


\bibitem[\protect\citeauthoryear{Arefin~Shimon, Lutton, Xu, Morrison-Smith,
  Boucher, and Ruiz}{Arefin~Shimon et~al\mbox{.}}{2016}]%
        {arefin_shimon_exploring_2016}
\bibfield{author}{\bibinfo{person}{Shaikh~Shawon Arefin~Shimon},
  \bibinfo{person}{Courtney Lutton}, \bibinfo{person}{Zichun Xu},
  \bibinfo{person}{Sarah Morrison-Smith}, \bibinfo{person}{Christina Boucher},
  {and} \bibinfo{person}{Jaime Ruiz}.} \bibinfo{year}{2016}\natexlab{}.
\newblock \showarticletitle{Exploring Non-touchscreen Gestures for
  Smartwatches}. In \bibinfo{booktitle}{\emph{Proceedings of the 2016 {CHI}
  Conference on Human Factors in Computing Systems}} (New York, {NY}, {USA})
  \emph{(\bibinfo{series}{{CHI} '16})}. \bibinfo{publisher}{Association for
  Computing Machinery}, \bibinfo{pages}{3822--3833}.
\newblock
\showISBNx{978-1-4503-3362-7}
\urldef\tempurl%
\url{https://doi.org/10.1145/2858036.2858385}
\showDOI{\tempurl}


\bibitem[\protect\citeauthoryear{Arif and Stuerzlinger}{Arif and
  Stuerzlinger}{2009}]%
        {arif_analysis_2009}
\bibfield{author}{\bibinfo{person}{Ahmed~Sabbir Arif} {and}
  \bibinfo{person}{Wolfgang Stuerzlinger}.} \bibinfo{year}{2009}\natexlab{}.
\newblock \showarticletitle{Analysis of text entry performance metrics}. In
  \bibinfo{booktitle}{\emph{2009 {IEEE} Toronto International Conference
  Science and Technology for Humanity ({TIC}-{STH})}} (Toronto, {ON}, Canada).
  \bibinfo{publisher}{{IEEE}}, \bibinfo{pages}{100--105}.
\newblock
\showISBNx{978-1-4244-3877-8}
\urldef\tempurl%
\url{https://doi.org/10.1109/TIC-STH.2009.5444533}
\showDOI{\tempurl}


\bibitem[\protect\citeauthoryear{Beaver, Wilson, and
  Schmitter-Edgecombe}{Beaver et~al\mbox{.}}{2019}]%
        {beaver_characterising_2019}
\bibfield{author}{\bibinfo{person}{Jenna Beaver}, \bibinfo{person}{Kaci~B.
  Wilson}, {and} \bibinfo{person}{Maureen Schmitter-Edgecombe}.}
  \bibinfo{year}{2019}\natexlab{}.
\newblock \showarticletitle{Characterising omission errors in everyday task
  completion and cognitive correlates in individuals with mild cognitive
  impairment and dementia}.
\newblock \bibinfo{journal}{\emph{Neuropsychological Rehabilitation}}
  \bibinfo{volume}{29}, \bibinfo{number}{5} (\bibinfo{date}{June}
  \bibinfo{year}{2019}), \bibinfo{pages}{804--820}.
\newblock
\showISSN{1464-0694}
\urldef\tempurl%
\url{https://doi.org/10.1080/09602011.2017.1337039}
\showDOI{\tempurl}


\bibitem[\protect\citeauthoryear{Bi and Zhai}{Bi and Zhai}{2016}]%
        {BiZhai16}
\bibfield{author}{\bibinfo{person}{Xiaojun Bi} {and} \bibinfo{person}{Shumin
  Zhai}.} \bibinfo{year}{2016}\natexlab{}.
\newblock \bibinfo{booktitle}{\emph{IJQwerty: What Difference Does One Key
  Change Make? Gesture Typing Keyboard Optimization Bounded by One Key Position
  Change from Qwerty}}.
\newblock \bibinfo{publisher}{Association for Computing Machinery},
  \bibinfo{address}{New York, NY, USA}, \bibinfo{pages}{49–58}.
\newblock
\showISBNx{9781450333627}
\urldef\tempurl%
\url{https://doi.org/10.1145/2858036.2858421}
\showURL{%
\tempurl}


\bibitem[\protect\citeauthoryear{Billah, Ko, Ashok, Bi, and
  Ramakrishnan}{Billah et~al\mbox{.}}{2019}]%
        {Bi12a}
\bibfield{author}{\bibinfo{person}{Syed~Masum Billah}, \bibinfo{person}{Yu-Jung
  Ko}, \bibinfo{person}{Vikas Ashok}, \bibinfo{person}{Xiaojun Bi}, {and}
  \bibinfo{person}{IV Ramakrishnan}.} \bibinfo{year}{2019}\natexlab{}.
\newblock \bibinfo{booktitle}{\emph{Accessible Gesture Typing for Non-Visual
  Text Entry on Smartphones}}.
\newblock \bibinfo{publisher}{Association for Computing Machinery},
  \bibinfo{address}{New York, NY, USA}, \bibinfo{pages}{1–12}.
\newblock
\showISBNx{9781450359702}
\urldef\tempurl%
\url{https://doi.org/10.1145/3290605.3300606}
\showURL{%
\tempurl}


\bibitem[\protect\citeauthoryear{Czaja}{Czaja}{2019}]%
        {czaja_usability_2019}
\bibfield{author}{\bibinfo{person}{Sara~J Czaja}.}
  \bibinfo{year}{2019}\natexlab{}.
\newblock \showarticletitle{Usability of {Technology} for {Older} {Adults}:
  {Where} {Are} {We} and {Where} {Do} {We} {Need} to {Be}}.
\newblock \bibinfo{journal}{\emph{Journal of Usability Studies}}
  \bibinfo{volume}{14}, \bibinfo{number}{2} (\bibinfo{year}{2019}),
  \bibinfo{pages}{4}.
\newblock


\bibitem[\protect\citeauthoryear{D'Arienzo}{D'Arienzo}{2022}]%
        {darienzo_keyoboard_2022}
\bibfield{author}{\bibinfo{person}{Armando D'Arienzo}.}
  \bibinfo{year}{2022}\natexlab{}.
\newblock \bibinfo{booktitle}{\emph{{KeyOboard}: a {WearOS} T9 Keyboard - Apps
  on Google Play}}.
\newblock
\urldef\tempurl%
\url{https://play.google.com/store/apps/details?id=com.armandodarienzo.wear.utility.KeyOboard&hl=en_US&gl=US}
\showURL{%
\tempurl}


\bibitem[\protect\citeauthoryear{Dolch}{Dolch}{1936}]%
        {Dolch1936}
\bibfield{author}{\bibinfo{person}{E.~W. Dolch}.}
  \bibinfo{year}{1936}\natexlab{}.
\newblock \showarticletitle{A Basic Sight Vocabulary}.
\newblock \bibinfo{journal}{\emph{The Elementary School Journal}}
  \bibinfo{volume}{36}, \bibinfo{number}{6} (\bibinfo{year}{1936}),
  \bibinfo{pages}{456--460}.
\newblock
\showISSN{00135984, 15548279}
\urldef\tempurl%
\url{http://www.jstor.org/stable/995914}
\showURL{%
\tempurl}


\bibitem[\protect\citeauthoryear{Dunlop and Masters}{Dunlop and
  Masters}{2008}]%
        {dunlop_investigating_2008}
\bibfield{author}{\bibinfo{person}{Mark Dunlop} {and} \bibinfo{person}{Michelle
  Masters}.} \bibinfo{year}{2008}\natexlab{}.
\newblock \showarticletitle{Investigating five key predictive text entry with
  combined distance and keystroke modelling}.
\newblock \bibinfo{journal}{\emph{Personal and Ubiquitous Computing}}
  \bibinfo{volume}{12} (\bibinfo{date}{Nov.} \bibinfo{year}{2008}),
  \bibinfo{pages}{589--598}.
\newblock
\urldef\tempurl%
\url{https://doi.org/10.1007/s00779-007-0179-7}
\showDOI{\tempurl}


\bibitem[\protect\citeauthoryear{Dunlop, Durga, Motaparti, Dona, and
  Medapuram}{Dunlop et~al\mbox{.}}{2012}]%
        {Dunlop12}
\bibfield{author}{\bibinfo{person}{Mark~D. Dunlop}, \bibinfo{person}{Naveen
  Durga}, \bibinfo{person}{Sunil Motaparti}, \bibinfo{person}{Prima Dona},
  {and} \bibinfo{person}{Varun Medapuram}.} \bibinfo{year}{2012}\natexlab{}.
\newblock \showarticletitle{QWERTH: An Optimized Semi-Ambiguous Keyboard
  Design}. In \bibinfo{booktitle}{\emph{Proceedings of the 14th International
  Conference on Human-Computer Interaction with Mobile Devices and Services
  Companion}} (San Francisco, California, USA)
  \emph{(\bibinfo{series}{MobileHCI '12})}. \bibinfo{publisher}{Association for
  Computing Machinery}, \bibinfo{address}{New York, NY, USA},
  \bibinfo{pages}{23–28}.
\newblock
\showISBNx{9781450314435}
\urldef\tempurl%
\url{https://doi.org/10.1145/2371664.2371671}
\showDOI{\tempurl}


\bibitem[\protect\citeauthoryear{Dunlop, Komninos, and Durga}{Dunlop
  et~al\mbox{.}}{2014}]%
        {dunlop_towards_2014}
\bibfield{author}{\bibinfo{person}{Mark~D. Dunlop}, \bibinfo{person}{Andreas
  Komninos}, {and} \bibinfo{person}{Naveen Durga}.}
  \bibinfo{year}{2014}\natexlab{}.
\newblock \showarticletitle{Towards high quality text entry on smartwatches}.
  In \bibinfo{booktitle}{\emph{{CHI} '14 {Extended} {Abstracts} on {Human}
  {Factors} in {Computing} {Systems}}} \emph{(\bibinfo{series}{{CHI} {EA}
  '14})}. \bibinfo{publisher}{Association for Computing Machinery},
  \bibinfo{address}{New York, NY, USA}, \bibinfo{pages}{2365--2370}.
\newblock
\showISBNx{978-1-4503-2474-8}
\urldef\tempurl%
\url{https://doi.org/10.1145/2559206.2581319}
\showDOI{\tempurl}


\bibitem[\protect\citeauthoryear{Fernández-Ardèvol and
  Rosales}{Fernández-Ardèvol and Rosales}{2017}]%
        {fernandez-ardevol_my_2017}
\bibfield{author}{\bibinfo{person}{Mireia Fernández-Ardèvol} {and}
  \bibinfo{person}{Andrea Rosales}.} \bibinfo{year}{2017}\natexlab{}.
\newblock \showarticletitle{My {Interests}, {My} {Activities}: {Learning} from
  an {Intergenerational} {Comparison} of {Smartwatch} {Use}}. In
  \bibinfo{booktitle}{\emph{Human {Aspects} of {IT} for the {Aged}
  {Population}. {Applications}, {Services} and {Contexts}}},
  \bibfield{editor}{\bibinfo{person}{Jia Zhou} {and} \bibinfo{person}{Gavriel
  Salvendy}} (Eds.). \bibinfo{publisher}{Springer International Publishing},
  \bibinfo{address}{Cham}, \bibinfo{pages}{114--129}.
\newblock
\showISBNx{978-3-319-58536-9}
\urldef\tempurl%
\url{https://doi.org/10.1007/978-3-319-58536-9_10}
\showDOI{\tempurl}


\bibitem[\protect\citeauthoryear{Fitts}{Fitts}{1954}]%
        {fitts_information_1954}
\bibfield{author}{\bibinfo{person}{Paul~M. Fitts}.}
  \bibinfo{year}{1954}\natexlab{}.
\newblock \showarticletitle{The information capacity of the human motor system
  in controlling the amplitude of movement}.
\newblock \bibinfo{journal}{\emph{Journal of Experimental Psychology}}
  \bibinfo{volume}{47}, \bibinfo{number}{6} (\bibinfo{year}{1954}),
  \bibinfo{pages}{381--391}.
\newblock
\showISSN{0022-1015}
\urldef\tempurl%
\url{https://doi.org/10.1037/h0055392}
\showDOI{\tempurl}
\newblock
\shownote{Place: US Publisher: American Psychological Association.}


\bibitem[\protect\citeauthoryear{Fuglerud, Chan, and Rli}{Fuglerud
  et~al\mbox{.}}{2018}]%
        {fuglerud_studying_2018}
\bibfield{author}{\bibinfo{person}{Kristin~S. Fuglerud},
  \bibinfo{person}{Richard Chan}, {and} \bibinfo{person}{Hilde~T. Rli}.}
  \bibinfo{year}{2018}\natexlab{}.
\newblock \showarticletitle{Studying {Older} {People} with {Visual}
  {Impairments} {Using} {Mainstream} {Smartphones} with the {Aid} of the
  {EziSmart} {Keypad} and {Apps}}.
\newblock \bibinfo{journal}{\emph{Transforming our World Through Design,
  Diversity and Education}} (\bibinfo{year}{2018}), \bibinfo{pages}{802--810}.
\newblock
\urldef\tempurl%
\url{https://doi.org/10.3233/978-1-61499-923-2-802}
\showDOI{\tempurl}
\newblock
\shownote{Publisher: IOS Press.}


\bibitem[\protect\citeauthoryear{Funk, Sahami, Henze, and Schmidt}{Funk
  et~al\mbox{.}}{2014}]%
        {funk_using_2014}
\bibfield{author}{\bibinfo{person}{Markus Funk}, \bibinfo{person}{Alireza
  Sahami}, \bibinfo{person}{Niels Henze}, {and} \bibinfo{person}{Albrecht
  Schmidt}.} \bibinfo{year}{2014}\natexlab{}.
\newblock \showarticletitle{Using a touch-sensitive wristband for text entry on
  smart watches}. In \bibinfo{booktitle}{\emph{{CHI} '14 Extended Abstracts on
  Human Factors in Computing Systems}} (New York, {NY}, {USA})
  \emph{(\bibinfo{series}{{CHI} {EA} '14})}. \bibinfo{publisher}{Association
  for Computing Machinery}, \bibinfo{pages}{2305--2310}.
\newblock
\showISBNx{978-1-4503-2474-8}
\urldef\tempurl%
\url{https://doi.org/10.1145/2559206.2581143}
\showDOI{\tempurl}


\bibitem[\protect\citeauthoryear{Google}{Google}{2022}]%
        {google_use_2022}
\bibfield{author}{\bibinfo{person}{Google}.} \bibinfo{year}{2022}\natexlab{}.
\newblock \bibinfo{title}{Use {Smart} {Compose} - {Computer} - {Gmail} {Help}}.
\newblock
\newblock
\urldef\tempurl%
\url{https://support.google.com/mail/answer/9116836?hl=en&co=GENIE.Platform\%3DDesktop}
\showURL{%
\tempurl}


\bibitem[\protect\citeauthoryear{Gordon, Ouyang, and Zhai}{Gordon
  et~al\mbox{.}}{2016}]%
        {Gordon_watchwriter_2016}
\bibfield{author}{\bibinfo{person}{Mitchell Gordon}, \bibinfo{person}{Tom
  Ouyang}, {and} \bibinfo{person}{Shumin Zhai}.}
  \bibinfo{year}{2016}\natexlab{}.
\newblock \showarticletitle{{WatchWriter}: Tap and Gesture Typing on a
  Smartwatch Miniature Keyboard with Statistical Decoding}. In
  \bibinfo{booktitle}{\emph{Proceedings of the 2016 {CHI} Conference on Human
  Factors in Computing Systems}} (San Jose California {USA}).
  \bibinfo{publisher}{{ACM}}, \bibinfo{pages}{3817--3821}.
\newblock
\showISBNx{978-1-4503-3362-7}
\urldef\tempurl%
\url{https://doi.org/10.1145/2858036.2858242}
\showDOI{\tempurl}


\bibitem[\protect\citeauthoryear{Hart and Staveland}{Hart and
  Staveland}{1988}]%
        {Hart1988nasa}
\bibfield{author}{\bibinfo{person}{Sandra~G. Hart} {and}
  \bibinfo{person}{Lowell~E. Staveland}.} \bibinfo{year}{1988}\natexlab{}.
\newblock \showarticletitle{Development of NASA-TLX (Task Load Index): Results
  of Empirical and Theoretical Research}.
\newblock In \bibinfo{booktitle}{\emph{Human Mental Workload}},
  \bibfield{editor}{\bibinfo{person}{Peter~A. Hancock} {and}
  \bibinfo{person}{Najmedin Meshkati}} (Eds.). \bibinfo{series}{Advances in
  Psychology}, Vol.~\bibinfo{volume}{52}. \bibinfo{publisher}{North-Holland},
  \bibinfo{pages}{139--183}.
\newblock
\showISSN{0166-4115}
\urldef\tempurl%
\url{https://doi.org/10.1016/S0166-4115(08)62386-9}
\showDOI{\tempurl}


\bibitem[\protect\citeauthoryear{Hawthorn}{Hawthorn}{2006}]%
        {hawthorn_designing_2006}
\bibfield{author}{\bibinfo{person}{Dan Hawthorn}.}
  \bibinfo{year}{2006}\natexlab{}.
\newblock \showarticletitle{Designing {Effective} {Interfaces} for {Older}
  {Users}}.
\newblock  (\bibinfo{date}{Jan.} \bibinfo{year}{2006}).
\newblock


\bibitem[\protect\citeauthoryear{Hisao}{Hisao}{1980}]%
        {hisao_historical_1980}
\bibfield{author}{\bibinfo{person}{Yamada Hisao}.}
  \bibinfo{year}{1980}\natexlab{}.
\newblock \showarticletitle{A Historical Study of Typewriters and Typing
  Methods: from the Position of Planning Japanese Parallels}.
\newblock  \bibinfo{volume}{2}, \bibinfo{number}{4} (\bibinfo{year}{1980}),
  \bibinfo{pages}{175--202}.
\newblock
\urldef\tempurl%
\url{https://cir.nii.ac.jp/crid/1050001337894287488}
\showURL{%
\tempurl}
\newblock
\shownote{Publisher: Information Processing Society of Japan ({IPSJ}).}


\bibitem[\protect\citeauthoryear{Hornbæk, Sander, Bargas-Avila, and
  Grue~Simonsen}{Hornbæk et~al\mbox{.}}{2014}]%
        {hornbaek_is_2014}
\bibfield{author}{\bibinfo{person}{Kasper Hornbæk}, \bibinfo{person}{Søren~S.
  Sander}, \bibinfo{person}{Javier~Andrés Bargas-Avila}, {and}
  \bibinfo{person}{Jakob Grue~Simonsen}.} \bibinfo{year}{2014}\natexlab{}.
\newblock \showarticletitle{Is once enough? on the extent and content of
  replications in human-computer interaction}. In
  \bibinfo{booktitle}{\emph{Proceedings of the {SIGCHI} Conference on Human
  Factors in Computing Systems}} (New York, {NY}, {USA})
  \emph{(\bibinfo{series}{{CHI} '14})}. \bibinfo{publisher}{Association for
  Computing Machinery}, \bibinfo{pages}{3523--3532}.
\newblock
\showISBNx{978-1-4503-2473-1}
\urldef\tempurl%
\url{https://doi.org/10.1145/2556288.2557004}
\showDOI{\tempurl}


\bibitem[\protect\citeauthoryear{Kalman, Kavé, and Umanski}{Kalman
  et~al\mbox{.}}{2015}]%
        {kalman_writing_2015}
\bibfield{author}{\bibinfo{person}{Yoram~M. Kalman}, \bibinfo{person}{Gitit
  Kavé}, {and} \bibinfo{person}{Daniil Umanski}.}
  \bibinfo{year}{2015}\natexlab{}.
\newblock \showarticletitle{Writing in a {Digital} {World}: {Self}-{Correction}
  {While} {Typing} in {Younger} and {Older} {Adults}}.
\newblock \bibinfo{journal}{\emph{International Journal of Environmental
  Research and Public Health}} \bibinfo{volume}{12}, \bibinfo{number}{10}
  (\bibinfo{date}{Oct.} \bibinfo{year}{2015}), \bibinfo{pages}{12723--12734}.
\newblock
\showISSN{1660-4601}
\urldef\tempurl%
\url{https://doi.org/10.3390/ijerph121012723}
\showDOI{\tempurl}
\newblock
\shownote{Number: 10 Publisher: Multidisciplinary Digital Publishing
  Institute.}


\bibitem[\protect\citeauthoryear{Kane, Wobbrock, Harniss, and Johnson}{Kane
  et~al\mbox{.}}{2008}]%
        {kane_truekeys_2008}
\bibfield{author}{\bibinfo{person}{Shaun~K. Kane}, \bibinfo{person}{Jacob~O.
  Wobbrock}, \bibinfo{person}{Mark Harniss}, {and} \bibinfo{person}{Kurt~L.
  Johnson}.} \bibinfo{year}{2008}\natexlab{}.
\newblock \showarticletitle{{TrueKeys}: identifying and correcting typing
  errors for people with motor impairments}. In
  \bibinfo{booktitle}{\emph{Proceedings of the 13th international conference on
  {Intelligent} user interfaces}} \emph{(\bibinfo{series}{{IUI} '08})}.
  \bibinfo{publisher}{Association for Computing Machinery},
  \bibinfo{address}{New York, NY, USA}, \bibinfo{pages}{349--352}.
\newblock
\showISBNx{978-1-59593-987-6}
\urldef\tempurl%
\url{https://doi.org/10.1145/1378773.1378827}
\showDOI{\tempurl}


\bibitem[\protect\citeauthoryear{Komninos and Dunlop}{Komninos and
  Dunlop}{2014}]%
        {komninos_text_2014}
\bibfield{author}{\bibinfo{person}{Andreas Komninos} {and}
  \bibinfo{person}{Mark Dunlop}.} \bibinfo{year}{2014}\natexlab{}.
\newblock \showarticletitle{Text {Input} on a {Smart} {Watch}}.
\newblock \bibinfo{journal}{\emph{IEEE Pervasive Computing}}
  \bibinfo{volume}{13}, \bibinfo{number}{4} (\bibinfo{date}{Oct.}
  \bibinfo{year}{2014}), \bibinfo{pages}{50--58}.
\newblock
\showISSN{1536-1268}
\urldef\tempurl%
\url{https://doi.org/10.1109/MPRV.2014.77}
\showDOI{\tempurl}


\bibitem[\protect\citeauthoryear{Kristensson}{Kristensson}{2009}]%
        {kristensson_five_2009}
\bibfield{author}{\bibinfo{person}{Per~Ola Kristensson}.}
  \bibinfo{year}{2009}\natexlab{}.
\newblock \showarticletitle{Five {Challenges} for {Intelligent} {Text} {Entry}
  {Methods}}.
\newblock \bibinfo{journal}{\emph{AI Magazine}} \bibinfo{volume}{30},
  \bibinfo{number}{4} (\bibinfo{date}{Sept.} \bibinfo{year}{2009}),
  \bibinfo{pages}{85--85}.
\newblock
\showISSN{2371-9621}
\urldef\tempurl%
\url{https://doi.org/10.1609/aimag.v30i4.2269}
\showDOI{\tempurl}
\newblock
\shownote{Number: 4.}


\bibitem[\protect\citeauthoryear{Kurniawan}{Kurniawan}{2008}]%
        {kurniawan_older_2008}
\bibfield{author}{\bibinfo{person}{Sri Kurniawan}.}
  \bibinfo{year}{2008}\natexlab{}.
\newblock \showarticletitle{Older people and mobile phones: {A} multi-method
  investigation}.
\newblock \bibinfo{journal}{\emph{International Journal of Human-Computer
  Studies}} \bibinfo{volume}{66}, \bibinfo{number}{12} (\bibinfo{date}{Dec.}
  \bibinfo{year}{2008}), \bibinfo{pages}{889--901}.
\newblock
\showISSN{1071-5819}
\urldef\tempurl%
\url{https://doi.org/10.1016/j.ijhcs.2008.03.002}
\showDOI{\tempurl}


\bibitem[\protect\citeauthoryear{Leiva, Sahami, Catala, Henze, and
  Schmidt}{Leiva et~al\mbox{.}}{2015a}]%
        {leiva_text_2015}
\bibfield{author}{\bibinfo{person}{Luis~A. Leiva}, \bibinfo{person}{Alireza
  Sahami}, \bibinfo{person}{Alejandro Catala}, \bibinfo{person}{Niels Henze},
  {and} \bibinfo{person}{Albrecht Schmidt}.} \bibinfo{year}{2015}\natexlab{a}.
\newblock \showarticletitle{Text Entry on Tiny {QWERTY} Soft Keyboards}. In
  \bibinfo{booktitle}{\emph{Proceedings of the 33rd Annual {ACM} Conference on
  Human Factors in Computing Systems}} (New York, {NY}, {USA})
  \emph{(\bibinfo{series}{{CHI} '15})}. \bibinfo{publisher}{Association for
  Computing Machinery}, \bibinfo{pages}{669--678}.
\newblock
\showISBNx{978-1-4503-3145-6}
\urldef\tempurl%
\url{https://doi.org/10.1145/2702123.2702388}
\showDOI{\tempurl}


\bibitem[\protect\citeauthoryear{Leiva, Sahami, Catalá, Henze, and
  Schmidt}{Leiva et~al\mbox{.}}{2015b}]%
        {leiva_error_2015}
\bibfield{author}{\bibinfo{person}{Luis~A Leiva}, \bibinfo{person}{Alireza
  Sahami}, \bibinfo{person}{Alejandro Catalá}, \bibinfo{person}{Niels Henze},
  {and} \bibinfo{person}{Albrecht Schmidt}.} \bibinfo{year}{2015}\natexlab{b}.
\newblock \showarticletitle{Error Auto-Correction Mechanisms on Tiny {QWERTY}
  Soft Keyboards}. In \bibinfo{booktitle}{\emph{CHI '15 Extended Abstracts on
  Human Factors in Computing Systems}} \emph{(\bibinfo{series}{{CHI} {EA}
  '15})}. \bibinfo{pages}{4}.
\newblock


\bibitem[\protect\citeauthoryear{Limited}{Limited}{2022}]%
        {vulcan_labs_company_limited_watchkey_2022}
\bibfield{author}{\bibinfo{person}{Vulcan Labs~Company Limited}.}
  \bibinfo{year}{2022}\natexlab{}.
\newblock \bibinfo{booktitle}{\emph{{WatchKey}: Keyboard for Watch}}.
\newblock
\urldef\tempurl%
\url{https://apps.apple.com/us/app/watchkey-keyboard-for-watch/id1499307793}
\showURL{%
\tempurl}


\bibitem[\protect\citeauthoryear{Lin, Zhu, Ko, Cui, and Bi}{Lin
  et~al\mbox{.}}{2018}]%
        {Lin18}
\bibfield{author}{\bibinfo{person}{Yu-Hao Lin}, \bibinfo{person}{Suwen Zhu},
  \bibinfo{person}{Yu-Jung Ko}, \bibinfo{person}{Wenzhe Cui}, {and}
  \bibinfo{person}{Xiaojun Bi}.} \bibinfo{year}{2018}\natexlab{}.
\newblock \showarticletitle{Why Is Gesture Typing Promising for Older Adults?
  Comparing Gesture and Tap Typing Behavior of Older with Young Adults}. In
  \bibinfo{booktitle}{\emph{Proceedings of the 20th International ACM SIGACCESS
  Conference on Computers and Accessibility}} (Galway, Ireland)
  \emph{(\bibinfo{series}{ASSETS '18})}. \bibinfo{publisher}{Association for
  Computing Machinery}, \bibinfo{address}{New York, NY, USA},
  \bibinfo{pages}{271–281}.
\newblock
\showISBNx{9781450356503}
\urldef\tempurl%
\url{https://doi.org/10.1145/3234695.3236350}
\showDOI{\tempurl}


\bibitem[\protect\citeauthoryear{MacKenzie}{MacKenzie}{2002}]%
        {Mac2002kspc}
\bibfield{author}{\bibinfo{person}{I.~Scott MacKenzie}.}
  \bibinfo{year}{2002}\natexlab{}.
\newblock \showarticletitle{KSPC (Keystrokes per Character) as a Characteristic
  of Text Entry Techniques}, In \bibinfo{booktitle}{Proceedings of the 4th
  International Symposium on Mobile Human-Computer Interaction}.
\newblock \bibinfo{journal}{\emph{Proc. Mobile HCI 2002}},
  \bibinfo{pages}{195--210}.
\newblock
\showISBNx{978-3-540-44189-2}
\urldef\tempurl%
\url{https://doi.org/10.1007/3-540-45756-9_16}
\showDOI{\tempurl}


\bibitem[\protect\citeauthoryear{MacKenzie}{MacKenzie}{2015}]%
        {Mac2015wpm}
\bibfield{author}{\bibinfo{person}{I.~Scott MacKenzie}.}
  \bibinfo{year}{2015}\natexlab{}.
\newblock \showarticletitle{A Note on Calculating Text Entry Speed}.
\newblock
\urldef\tempurl%
\url{https://www.yorku.ca/mack/RN-TextEntrySpeed.html}
\showURL{%
\tempurl}


\bibitem[\protect\citeauthoryear{MacKenzie and Soukoreff}{MacKenzie and
  Soukoreff}{2003}]%
        {Mac2003phrase}
\bibfield{author}{\bibinfo{person}{I.~Scott MacKenzie} {and}
  \bibinfo{person}{R.~William Soukoreff}.} \bibinfo{year}{2003}\natexlab{}.
\newblock \showarticletitle{Phrase Sets for Evaluating Text Entry Techniques}.
  In \bibinfo{booktitle}{\emph{CHI '03 Extended Abstracts on Human Factors in
  Computing Systems}} (Ft. Lauderdale, Florida, USA)
  \emph{(\bibinfo{series}{CHI EA '03})}. \bibinfo{publisher}{Association for
  Computing Machinery}, \bibinfo{address}{New York, NY, USA},
  \bibinfo{pages}{754–755}.
\newblock
\showISBNx{1581136374}
\urldef\tempurl%
\url{https://doi.org/10.1145/765891.765971}
\showDOI{\tempurl}


\bibitem[\protect\citeauthoryear{Maiti, Jadliwala, and Weber}{Maiti
  et~al\mbox{.}}{2017}]%
        {maiti_preventing_2017}
\bibfield{author}{\bibinfo{person}{Anindya Maiti}, \bibinfo{person}{Murtuza
  Jadliwala}, {and} \bibinfo{person}{Chase Weber}.}
  \bibinfo{year}{2017}\natexlab{}.
\newblock \showarticletitle{Preventing shoulder surfing using randomized
  augmented reality keyboards}. In \bibinfo{booktitle}{\emph{2017 {IEEE}
  International Conference on Pervasive Computing and Communications Workshops
  ({PerCom} Workshops)}}. \bibinfo{pages}{630--635}.
\newblock
\urldef\tempurl%
\url{https://doi.org/10.1109/PERCOMW.2017.7917636}
\showDOI{\tempurl}


\bibitem[\protect\citeauthoryear{Manini, Mendoza, Battula, Davoudi,
  Kheirkhahan, Young, Weber, Fillingim, and Rashidi}{Manini
  et~al\mbox{.}}{2019}]%
        {manini_perception_2019}
\bibfield{author}{\bibinfo{person}{Todd~Matthew Manini},
  \bibinfo{person}{Tonatiuh Mendoza}, \bibinfo{person}{Manoj Battula},
  \bibinfo{person}{Anis Davoudi}, \bibinfo{person}{Matin Kheirkhahan},
  \bibinfo{person}{Mary~Ellen Young}, \bibinfo{person}{Eric Weber},
  \bibinfo{person}{Roger~Benton Fillingim}, {and} \bibinfo{person}{Parisa
  Rashidi}.} \bibinfo{year}{2019}\natexlab{}.
\newblock \showarticletitle{Perception of {Older} {Adults} {Toward}
  {Smartwatch} {Technology} for {Assessing} {Pain} and {Related}
  {Patient}-{Reported} {Outcomes}: {Pilot} {Study}}.
\newblock \bibinfo{journal}{\emph{JMIR mHealth and uHealth}}
  \bibinfo{volume}{7}, \bibinfo{number}{3} (\bibinfo{date}{March}
  \bibinfo{year}{2019}), \bibinfo{pages}{e10044}.
\newblock
\urldef\tempurl%
\url{https://doi.org/10.2196/10044}
\showDOI{\tempurl}
\newblock
\shownote{Company: JMIR mHealth and uHealth Distributor: JMIR mHealth and
  uHealth Institution: JMIR mHealth and uHealth Label: JMIR mHealth and uHealth
  Publisher: JMIR Publications Inc., Toronto, Canada.}


\bibitem[\protect\citeauthoryear{Microsoft}{Microsoft}{2022}]%
        {microsoft_enable_2022}
\bibfield{author}{\bibinfo{person}{Microsoft}.}
  \bibinfo{year}{2022}\natexlab{}.
\newblock \bibinfo{title}{Enable text suggestions for inclusive classrooms}.
\newblock
\newblock
\urldef\tempurl%
\url{https://support.microsoft.com/en-us/topic/enable-text-suggestions-for-inclusive-classrooms-8ca3ea32-66b1-4d9d-abf1-1d0ead34f2a2}
\showURL{%
\tempurl}


\bibitem[\protect\citeauthoryear{Nations}{Nations}{[n.d.]}]%
        {united_nations_older_2022}
\bibfield{author}{\bibinfo{person}{United Nations}.}
  \bibinfo{year}{[n.d.]}\natexlab{}.
\newblock \bibinfo{booktitle}{\emph{Older persons - {UNHCR}{\textbar}Emergency
  Handbook}}.
\newblock
\urldef\tempurl%
\url{https://emergency.unhcr.org/entry/43935/older-persons}
\showURL{%
\tempurl}


\bibitem[\protect\citeauthoryear{Nicolau and Jorge}{Nicolau and Jorge}{2012}]%
        {nicolau_elderly_2012}
\bibfield{author}{\bibinfo{person}{Hugo Nicolau} {and} \bibinfo{person}{Joaquim
  Jorge}.} \bibinfo{year}{2012}\natexlab{}.
\newblock \showarticletitle{Elderly text-entry performance on touchscreens}. In
  \bibinfo{booktitle}{\emph{Proceedings of the 14th international {ACM}
  {SIGACCESS} conference on {Computers} and accessibility}}
  \emph{(\bibinfo{series}{{ASSETS} '12})}. \bibinfo{publisher}{Association for
  Computing Machinery}, \bibinfo{address}{New York, NY, USA},
  \bibinfo{pages}{127--134}.
\newblock
\showISBNx{978-1-4503-1321-6}
\urldef\tempurl%
\url{https://doi.org/10.1145/2384916.2384939}
\showDOI{\tempurl}


\bibitem[\protect\citeauthoryear{Nielsen}{Nielsen}{1994}]%
        {nielsen_10_1994}
\bibfield{author}{\bibinfo{person}{Jakob Nielsen}.}
  \bibinfo{year}{1994}\natexlab{}.
\newblock \bibinfo{title}{10 {Usability} {Heuristics} for {User} {Interface}
  {Design}}.
\newblock
\newblock
\urldef\tempurl%
\url{https://www.nngroup.com/articles/ten-usability-heuristics/}
\showURL{%
\tempurl}


\bibitem[\protect\citeauthoryear{Oney, Harrison, Ogan, and Wiese}{Oney
  et~al\mbox{.}}{2013}]%
        {oney_zoomboard_2013}
\bibfield{author}{\bibinfo{person}{Stephen Oney}, \bibinfo{person}{Chris
  Harrison}, \bibinfo{person}{Amy Ogan}, {and} \bibinfo{person}{Jason Wiese}.}
  \bibinfo{year}{2013}\natexlab{}.
\newblock \showarticletitle{{ZoomBoard}: a diminutive qwerty soft keyboard
  using iterative zooming for ultra-small devices}. In
  \bibinfo{booktitle}{\emph{Proceedings of the {SIGCHI} Conference on Human
  Factors in Computing Systems}} (New York, {NY}, {USA})
  \emph{(\bibinfo{series}{{CHI} '13})}. \bibinfo{publisher}{Association for
  Computing Machinery}, \bibinfo{pages}{2799--2802}.
\newblock
\showISBNx{978-1-4503-1899-0}
\urldef\tempurl%
\url{https://doi.org/10.1145/2470654.2481387}
\showDOI{\tempurl}


\bibitem[\protect\citeauthoryear{Organization}{Organization}{2021}]%
        {world_health_organization_ageing_2021}
\bibfield{author}{\bibinfo{person}{World~Health Organization}.}
  \bibinfo{year}{2021}\natexlab{}.
\newblock \bibinfo{title}{Ageing and health}.
\newblock
\newblock
\urldef\tempurl%
\url{https://www.who.int/news-room/fact-sheets/detail/ageing-and-health}
\showURL{%
\tempurl}


\bibitem[\protect\citeauthoryear{Palin, Feit, Kim, Kristensson, and
  Oulasvirta}{Palin et~al\mbox{.}}{2019}]%
        {Palin19}
\bibfield{author}{\bibinfo{person}{Kseniia Palin}, \bibinfo{person}{Anna~Maria
  Feit}, \bibinfo{person}{Sunjun Kim}, \bibinfo{person}{Per~Ola Kristensson},
  {and} \bibinfo{person}{Antti Oulasvirta}.} \bibinfo{year}{2019}\natexlab{}.
\newblock \showarticletitle{How Do People Type on Mobile Devices? Observations
  from a Study with 37,000 Volunteers}. In
  \bibinfo{booktitle}{\emph{Proceedings of the 21st International Conference on
  Human-Computer Interaction with Mobile Devices and Services}} (Taipei,
  Taiwan) \emph{(\bibinfo{series}{MobileHCI '19})}.
  \bibinfo{publisher}{Association for Computing Machinery},
  \bibinfo{address}{New York, NY, USA}, Article \bibinfo{articleno}{9},
  \bibinfo{numpages}{12}~pages.
\newblock
\showISBNx{9781450368254}
\urldef\tempurl%
\url{https://doi.org/10.1145/3338286.3340120}
\showDOI{\tempurl}


\bibitem[\protect\citeauthoryear{Perrault, Lecolinet, Eagan, and
  Guiard}{Perrault et~al\mbox{.}}{2013}]%
        {perrault_watchit_2013}
\bibfield{author}{\bibinfo{person}{Simon~T. Perrault}, \bibinfo{person}{Eric
  Lecolinet}, \bibinfo{person}{James Eagan}, {and} \bibinfo{person}{Yves
  Guiard}.} \bibinfo{year}{2013}\natexlab{}.
\newblock \showarticletitle{Watchit: simple gestures and eyes-free interaction
  for wristwatches and bracelets}. In \bibinfo{booktitle}{\emph{Proceedings of
  the {SIGCHI} Conference on Human Factors in Computing Systems}} (Paris
  France). \bibinfo{publisher}{{ACM}}, \bibinfo{pages}{1451--1460}.
\newblock
\showISBNx{978-1-4503-1899-0}
\urldef\tempurl%
\url{https://doi.org/10.1145/2470654.2466192}
\showDOI{\tempurl}


\bibitem[\protect\citeauthoryear{Porsager}{Porsager}{2022}]%
        {porsager_type_2022}
\bibfield{author}{\bibinfo{person}{Porsager}.} \bibinfo{year}{2022}\natexlab{}.
\newblock \bibinfo{booktitle}{\emph{Type Nine - T9 Keyboard}}.
\newblock
\urldef\tempurl%
\url{https://apps.apple.com/ca/app/type-nine-t9-keyboard/id926008509}
\showURL{%
\tempurl}


\bibitem[\protect\citeauthoryear{Qin, Zhu, Lin, Ko, and Bi}{Qin
  et~al\mbox{.}}{2018}]%
        {Qin18}
\bibfield{author}{\bibinfo{person}{Ryan Qin}, \bibinfo{person}{Suwen Zhu},
  \bibinfo{person}{Yu-Hao Lin}, \bibinfo{person}{Yu-Jung Ko}, {and}
  \bibinfo{person}{Xiaojun Bi}.} \bibinfo{year}{2018}\natexlab{}.
\newblock \showarticletitle{Optimal-T9: An Optimized T9-like Keyboard for Small
  Touchscreen Devices}. In \bibinfo{booktitle}{\emph{Proceedings of the 2018
  ACM International Conference on Interactive Surfaces and Spaces}} (Tokyo,
  Japan) \emph{(\bibinfo{series}{ISS '18})}. \bibinfo{publisher}{Association
  for Computing Machinery}, \bibinfo{address}{New York, NY, USA},
  \bibinfo{pages}{137–146}.
\newblock
\showISBNx{9781450356947}
\urldef\tempurl%
\url{https://doi.org/10.1145/3279778.3279786}
\showDOI{\tempurl}


\bibitem[\protect\citeauthoryear{Romero}{Romero}{2022}]%
        {romero_how_2022}
\bibfield{author}{\bibinfo{person}{Andrew Romero}.}
  \bibinfo{year}{2022}\natexlab{}.
\newblock \bibinfo{booktitle}{\emph{How to switch to a full {QWERTY} keyboard
  on your Galaxy Watch}}.
\newblock
\urldef\tempurl%
\url{https://9to5google.com/2022/07/12/qwerty-keyboard-on-galaxy-watch/}
\showURL{%
\tempurl}


\bibitem[\protect\citeauthoryear{Sarcar}{Sarcar}{2019}]%
        {sarcar_ability-based_2019}
\bibfield{author}{\bibinfo{person}{Sayan Sarcar}.}
  \bibinfo{year}{2019}\natexlab{}.
\newblock \showarticletitle{Ability-based {Optimization}: {Design} and
  {Evaluation} of {Touchscreen} {Keyboards} for {Older} {Adults} with
  {Dyslexia}}. In \bibinfo{booktitle}{\emph{Proceedings of the 31st
  {Australian} {Conference} on {Human}-{Computer}-{Interaction}}}
  \emph{(\bibinfo{series}{{OZCHI}'19})}. \bibinfo{publisher}{Association for
  Computing Machinery}, \bibinfo{address}{New York, NY, USA},
  \bibinfo{pages}{472--475}.
\newblock
\showISBNx{978-1-4503-7696-9}
\urldef\tempurl%
\url{https://doi.org/10.1145/3369457.3369519}
\showDOI{\tempurl}


\bibitem[\protect\citeauthoryear{Sarcar, Jokinen, Oulasvirta, Wang,
  Silpasuwanchai, and Ren}{Sarcar et~al\mbox{.}}{2018}]%
        {sarcar_ability-based_2018}
\bibfield{author}{\bibinfo{person}{Sayan Sarcar}, \bibinfo{person}{Jussi P.~P.
  Jokinen}, \bibinfo{person}{Antti Oulasvirta}, \bibinfo{person}{Zhenxin Wang},
  \bibinfo{person}{Chaklam Silpasuwanchai}, {and} \bibinfo{person}{Xiangshi
  Ren}.} \bibinfo{year}{2018}\natexlab{}.
\newblock \showarticletitle{Ability-{Based} {Optimization} of {Touchscreen}
  {Interactions}}.
\newblock \bibinfo{journal}{\emph{IEEE Pervasive Computing}}
  \bibinfo{volume}{17}, \bibinfo{number}{01} (\bibinfo{date}{Jan.}
  \bibinfo{year}{2018}), \bibinfo{pages}{15--26}.
\newblock
\showISSN{1536-1268}
\urldef\tempurl%
\url{https://doi.org/10.1109/MPRV.2018.011591058}
\showDOI{\tempurl}
\newblock
\shownote{Publisher: IEEE Computer Society.}


\bibitem[\protect\citeauthoryear{Seifert}{Seifert}{2020}]%
        {seifert_smartwatch_2020}
\bibfield{author}{\bibinfo{person}{Alexander Seifert}.}
  \bibinfo{year}{2020}\natexlab{}.
\newblock \showarticletitle{Smartwatch {Use} {Among} {Older} {Adults}:
  {Findings} from {Two} {Large} {Surveys}}. In \bibinfo{booktitle}{\emph{Human
  {Aspects} of {IT} for the {Aged} {Population}. {Technologies}, {Design} and
  {User} {Experience}}}, \bibfield{editor}{\bibinfo{person}{Qin Gao} {and}
  \bibinfo{person}{Jia Zhou}} (Eds.). \bibinfo{publisher}{Springer
  International Publishing}, \bibinfo{address}{Cham},
  \bibinfo{pages}{372--385}.
\newblock
\showISBNx{978-3-030-50252-2}
\urldef\tempurl%
\url{https://doi.org/10.1007/978-3-030-50252-2_28}
\showDOI{\tempurl}


\bibitem[\protect\citeauthoryear{Siek, Rogers, and Connelly}{Siek
  et~al\mbox{.}}{2005}]%
        {Siek2005fatfinger}
\bibfield{author}{\bibinfo{person}{Katie~A. Siek}, \bibinfo{person}{Yvonne
  Rogers}, {and} \bibinfo{person}{Kay~H. Connelly}.}
  \bibinfo{year}{2005}\natexlab{}.
\newblock \showarticletitle{Fat Finger Worries: How Older and Younger Users
  Physically Interact with {PDAs}}.
\newblock In \bibinfo{booktitle}{\emph{Human-Computer Interaction - {INTERACT}
  2005}}. \bibinfo{publisher}{Springer Berlin Heidelberg},
  \bibinfo{pages}{267--280}.
\newblock
\urldef\tempurl%
\url{https://doi.org/10.1007/11555261_24}
\showDOI{\tempurl}


\bibitem[\protect\citeauthoryear{Smith, Bi, and Zhai}{Smith
  et~al\mbox{.}}{2015}]%
        {Smith15}
\bibfield{author}{\bibinfo{person}{Brian~A. Smith}, \bibinfo{person}{Xiaojun
  Bi}, {and} \bibinfo{person}{Shumin Zhai}.} \bibinfo{year}{2015}\natexlab{}.
\newblock \bibinfo{booktitle}{\emph{Optimizing Touchscreen Keyboards for
  Gesture Typing}}.
\newblock \bibinfo{publisher}{Association for Computing Machinery},
  \bibinfo{address}{New York, NY, USA}, \bibinfo{pages}{3365–3374}.
\newblock
\showISBNx{9781450331456}
\urldef\tempurl%
\url{https://doi.org/10.1145/2702123.2702357}
\showURL{%
\tempurl}


\bibitem[\protect\citeauthoryear{Soukoreff and {MacKenzie}}{Soukoreff and
  {MacKenzie}}{2003}]%
        {soukoreff_metrics_2003}
\bibfield{author}{\bibinfo{person}{R.~William Soukoreff} {and}
  \bibinfo{person}{I.~Scott {MacKenzie}}.} \bibinfo{year}{2003}\natexlab{}.
\newblock \showarticletitle{Metrics for text entry research: an evaluation of
  {MSD} and {KSPC}, and a new unified error metric}. In
  \bibinfo{booktitle}{\emph{Proceedings of the {SIGCHI} Conference on Human
  Factors in Computing Systems}} (New York, {NY}, {USA})
  \emph{(\bibinfo{series}{{CHI} '03})}. \bibinfo{publisher}{Association for
  Computing Machinery}, \bibinfo{pages}{113--120}.
\newblock
\showISBNx{978-1-58113-630-2}
\urldef\tempurl%
\url{https://doi.org/10.1145/642611.642632}
\showDOI{\tempurl}


\bibitem[\protect\citeauthoryear{{SPSS}}{{SPSS}}{2022}]%
        {spss_effect_2022}
\bibfield{author}{\bibinfo{person}{{SPSS}}.} \bibinfo{year}{2022}\natexlab{}.
\newblock \bibinfo{booktitle}{\emph{Effect Size in Statistics - The Ultimate
  Guide}}.
\newblock
\urldef\tempurl%
\url{https://www.spss-tutorials.com/effect-size/}
\showURL{%
\tempurl}


\bibitem[\protect\citeauthoryear{Stremmel and Singh}{Stremmel and
  Singh}{2021}]%
        {stremmel_pretraining_2021}
\bibfield{author}{\bibinfo{person}{Joel Stremmel} {and} \bibinfo{person}{Arjun
  Singh}.} \bibinfo{year}{2021}\natexlab{}.
\newblock \showarticletitle{Pretraining {Federated} {Text} {Models} for {Next}
  {Word} {Prediction}}. In \bibinfo{booktitle}{\emph{Advances in {Information}
  and {Communication}}}, \bibfield{editor}{\bibinfo{person}{Kohei Arai}} (Ed.).
  \bibinfo{publisher}{Springer International Publishing},
  \bibinfo{address}{Cham}, \bibinfo{pages}{477--488}.
\newblock
\showISBNx{978-3-030-73103-8}
\urldef\tempurl%
\url{https://doi.org/10.1007/978-3-030-73103-8_34}
\showDOI{\tempurl}


\bibitem[\protect\citeauthoryear{Vertanen, Memmi, Emge, Reyal, and
  Kristensson}{Vertanen et~al\mbox{.}}{2015}]%
        {Vertanen_velocitap_2015}
\bibfield{author}{\bibinfo{person}{Keith Vertanen}, \bibinfo{person}{Haythem
  Memmi}, \bibinfo{person}{Justin Emge}, \bibinfo{person}{Shyam Reyal}, {and}
  \bibinfo{person}{Per~Ola Kristensson}.} \bibinfo{year}{2015}\natexlab{}.
\newblock \showarticletitle{{VelociTap}: Investigating Fast Mobile Text Entry
  using Sentence-Based Decoding of Touchscreen Keyboard Input}. In
  \bibinfo{booktitle}{\emph{Proceedings of the 33rd Annual {ACM} Conference on
  Human Factors in Computing Systems}} (New York, {NY}, {USA})
  \emph{(\bibinfo{series}{{CHI} '15})}. \bibinfo{publisher}{Association for
  Computing Machinery}, \bibinfo{pages}{659--668}.
\newblock
\showISBNx{978-1-4503-3145-6}
\urldef\tempurl%
\url{https://doi.org/10.1145/2702123.2702135}
\showDOI{\tempurl}


\bibitem[\protect\citeauthoryear{Voelcker-Rehage}{Voelcker-Rehage}{2008}]%
        {voelcker-rehage_motor-skill_2008}
\bibfield{author}{\bibinfo{person}{Claudia Voelcker-Rehage}.}
  \bibinfo{year}{2008}\natexlab{}.
\newblock \showarticletitle{Motor-skill learning in older adults—a review of
  studies on age-related differences}.
\newblock \bibinfo{journal}{\emph{European Review of Aging and Physical
  Activity}} \bibinfo{volume}{5}, \bibinfo{number}{1} (\bibinfo{date}{April}
  \bibinfo{year}{2008}), \bibinfo{pages}{5--16}.
\newblock
\showISSN{1861-6909}
\urldef\tempurl%
\url{https://doi.org/10.1007/s11556-008-0030-9}
\showDOI{\tempurl}
\newblock
\shownote{Number: 1 Publisher: BioMed Central.}


\bibitem[\protect\citeauthoryear{Wobbrock and Myers}{Wobbrock and
  Myers}{2006}]%
        {wobbrock_analyzing_2006}
\bibfield{author}{\bibinfo{person}{Jacob Wobbrock} {and} \bibinfo{person}{Brad
  Myers}.} \bibinfo{year}{2006}\natexlab{}.
\newblock \showarticletitle{Analyzing the input stream for character- level
  errors in unconstrained text entry evaluations}.
\newblock \bibinfo{journal}{\emph{ACM Trans. Comput.-Hum. Interact.}}
  \bibinfo{volume}{13} (\bibinfo{date}{Dec.} \bibinfo{year}{2006}),
  \bibinfo{pages}{458--489}.
\newblock
\urldef\tempurl%
\url{https://doi.org/10.1145/1188816.1188819}
\showDOI{\tempurl}


\bibitem[\protect\citeauthoryear{Wobbrock, Kane, Gajos, Harada, and
  Froehlich}{Wobbrock et~al\mbox{.}}{2011}]%
        {wobbrock_ability-based_2011}
\bibfield{author}{\bibinfo{person}{Jacob~O. Wobbrock},
  \bibinfo{person}{Shaun~K. Kane}, \bibinfo{person}{Krzysztof~Z. Gajos},
  \bibinfo{person}{Susumu Harada}, {and} \bibinfo{person}{Jon Froehlich}.}
  \bibinfo{year}{2011}\natexlab{}.
\newblock \showarticletitle{Ability-{Based} {Design}: {Concept}, {Principles}
  and {Examples}}.
\newblock \bibinfo{journal}{\emph{ACM Transactions on Accessible Computing}}
  \bibinfo{volume}{3}, \bibinfo{number}{3} (\bibinfo{date}{April}
  \bibinfo{year}{2011}), \bibinfo{pages}{9:1--9:27}.
\newblock
\showISSN{1936-7228}
\urldef\tempurl%
\url{https://doi.org/10.1145/1952383.1952384}
\showDOI{\tempurl}


\bibitem[\protect\citeauthoryear{Wong, Zhu, and Fu}{Wong et~al\mbox{.}}{2018}]%
        {Wong2018fingert9}
\bibfield{author}{\bibinfo{person}{Pui~Chung Wong}, \bibinfo{person}{Kening
  Zhu}, {and} \bibinfo{person}{Hongbo Fu}.} \bibinfo{year}{2018}\natexlab{}.
\newblock \bibinfo{booktitle}{\emph{FingerT9: Leveraging Thumb-to-Finger
  Interaction for Same-Side-Hand Text Entry on Smartwatches}}.
\newblock \bibinfo{publisher}{Association for Computing Machinery},
  \bibinfo{address}{New York, NY, USA}, \bibinfo{pages}{1–10}.
\newblock
\showISBNx{9781450356206}
\urldef\tempurl%
\url{https://doi.org/10.1145/3173574.3173752}
\showURL{%
\tempurl}


\bibitem[\protect\citeauthoryear{Zhai and Kristensson}{Zhai and
  Kristensson}{2012}]%
        {Zhai12}
\bibfield{author}{\bibinfo{person}{Shumin Zhai} {and} \bibinfo{person}{Per~Ola
  Kristensson}.} \bibinfo{year}{2012}\natexlab{}.
\newblock \showarticletitle{The Word-Gesture Keyboard: Reimagining Keyboard
  Interaction}.
\newblock \bibinfo{journal}{\emph{Commun. ACM}} \bibinfo{volume}{55},
  \bibinfo{number}{9} (\bibinfo{date}{Sept.} \bibinfo{year}{2012}),
  \bibinfo{pages}{91–101}.
\newblock
\showISSN{0001-0782}
\urldef\tempurl%
\url{https://doi.org/10.1145/2330667.2330689}
\showDOI{\tempurl}


\end{thebibliography}


\end{document}